\documentclass[useAMS,usenatbib,usegraphicx]{mn2e}

\usepackage{color}

\title[The MARCOS tool for DSA]{MARCOS, a numerical tool for the simulation of multiple time-dependent non-linear diffusive shock acceleration}
\author[Ferrand Downes Marcowith]{Gilles Ferrand$^{1}$, Turlough Downes$^{2}$, Alexandre Marcowith$^{3}$\\
$^{1}$CESR, Universit{\'e} Paul Sabatier, CNRS, 9 avenue du Colonel Roche, 31028 Toulouse, France\\
$^{2}$School of Mathematical Science, Dublin City University, Glasnevin, Dublin 9, Ireland\\
$^{3}$LPTA, Universit{\'e} Montpellier II, CNRS, B{\^a}t. 13, place Eug{\`e}ne Bataillon, 34095 Montpellier, France}

\begin{document}

\date{Accepted ?. Received ?; in original form ?}

\pagerange{\pageref{firstpage}--\pageref{lastpage}} \pubyear{2007}

\maketitle

\label{firstpage}

\begin{abstract}
We present a new code aimed at the simulation of diffusive shock acceleration (DSA), and discuss various test cases which demonstrate its ability to study DSA in its full time-dependent and non-linear developments. We present the numerical methods implemented, coupling the hydrodynamical evolution of a parallel shock (in one space dimension) and the kinetic transport of the cosmic-rays (CR) distribution function (in one momentum dimension), as first done by Falle. Following Kang and Jones and collaborators, we show how the \emph{adaptive mesh refinement} technique (AMR) greatly helps accommodating the extremely demanding numerical resolution requirements of realistic (Bohm-like) CR diffusion coefficients. We also present the \emph{parallelization} of the code, which allows us to run many successive shocks at the cost of a single shock, and thus to present the first direct numerical simulations of linear and non-linear \emph{multiple} DSA, a mechanism of interest in various astrophysical environments such as superbubbles, galaxy clusters and early cosmological flows.
\end{abstract}

\begin{keywords}
acceleration of particles -- methods~: numerical
\end{keywords}

\section{Introduction}

Diffusive Shock Acceleration (DSA) at supernova remnant blast waves is the favoured production mechanism for the production of the galactic cosmic-rays (CR). This theory, developed since the late 70s (see \citealt{Drury1983a} for a review), has now both strong theoretical and observational supports. The theoretical grounds of the model lie in the early ideas of Fermi (\citeyear{Fermi1949a}, \citeyear{Fermi1954a}): the regular Fermi acceleration mechanism (known as Fermi~I, the stochastic one being known as Fermi~II) can naturally explain the formation of a power-law spectrum by a shock wave -- with a remarkable universal slope whose value $s$ depends solely on the shock compression ratio $r$ (which is always 4 for strong non-relativistic shocks). However, the acceleration process can easily be so efficient that the CR may back-react on the shock dynamics, modifying the acceleration process in a fully non-linear way, and requiring a much more detailed analysis (see \citealt{Malkov2001c} for a review). 

Thus the DSA mechanism has still received a lot of attention in the last 20 years, from both a theoretical and a numerical perspective. Analytical works have been mostly limited to the test-particle (linear) regime. The full non-linear time-dependent problem has been mostly investigated through numerical simulations, using several different approaches (see \citealt{Jones2001a} for a short review). A first class is based on particle methods, from the early Monte-Carlo simulations developed by \cite{Ellison1984a} to the recent Particle-In-Cells codes (eg \citealt{Dieckmann2000a}). An alternate approach consists of solving the (Fokker-Planck) transport equation. This has first been done in the "two fluid" model (eg \citealt{Jones1990a}), then dealing with the complete particle distribution function (\citealt{Falle1987a}, \citealt{Bell1987a}, \citealt{Kang1991a}, \citealt{Duffy1992a}).

Most of this work has been aimed at understanding the role of single (isolated) supernovae remnants, although in many contexts CR are likely to experience many shocks, most notably inside superbubbles (\citealt{Parizot2004a}). 

In this paper we present a new code for the study of DSA, named \emph{Marcos} for \emph{Machine {\`a} Acc{\'e}l{\'e}rer les Rayons COSmiques}\footnote{the French for \emph{COSmic-Rays Acceleration Machine}}. In section~\ref{sec-dsa} we present the basics of the numerical methods implemented in our code, which couples the hydrodynamical evolution of a fluid with the kinetic transport of the CR. In section~\ref{sec-scales} we present the Adaptive Mesh Refinement (AMR) technique which allows us to resolve the (very) different scales induced by CR diffusion. In section~\ref{sec-multi} we present parallelization of the code, to be able to study in reasonable wall-clock time the effects of multiple shocks.

\section{Accelerating particles: coupling hydrodynamic and kinetic transports}
\label{sec-dsa}

In the DSA mechanism particles can be accelerated up to very high energies because they are interacting a large number of times with a macroscopic structure, namely a shock wave (as the forward shock of a supernova remnant). In our approach the thermal (fluid being shocked) and non-thermal (CR) particles, although intimately coupled, are handled as two different populations. The fluid, described by its moments $\rho$, $u$, $P$, obeys the Euler equations, while the CR, described by their distribution function $f(x,p)$, follow a more general transport equation.

\subsection{Hydrodynamics}
\label{sec-dsa-hydro}

The hydrodynamics are described by the (non-relativistic) Euler equations, which express the conservation of the fluid mass density $\rho$, momentum $\rho \vec{u}$ and total energy density $e$ and can be written in the general form:
\begin{equation}
\label{eq-cons}
\frac{\partial\vec{X}}{\partial{t}} + \vec{\nabla}\cdot\vec{F}\left(\vec{X}\right) = 0
\end{equation}
where (from now on in 1D slab geometry\footnote{
For the sake of simplicity the code is presented here in 1D slab geometry, but it works in 1D symmetrical spherical geometry too, provided we take into account the correct geometry of the finite-volume cells, which are now shells. The formalism of section~\ref{sec-dsa-hydro} still holds with $u$ being the radial velocity in equations (\ref{eq-cons-var}) and (\ref{eq-cons-flux}) and with a source term
\[
\vec{Q} = \left( \begin{array}{c} 0 \\ 2P/x \\ 0 \end{array} \right)
\]
added to equation~(\ref{eq-cons}).
}) the conservative variables are
\begin{equation}
\label{eq-cons-var}
\vec{X} = \left( \begin{array}{c} \rho \\ \rho{u} \\ e \end{array} \right)
\end{equation}
and their flux
\begin{equation}
\label{eq-cons-flux}
\vec{F}\left(\vec{X}\right) = \left( \begin{array}{c} \rho{u} \\ \rho{u^2}+P \\ (e+P)u \end{array} \right).
\end{equation}

To close this system we consider the usual polytropic equation of state of "adiabatic index" $\gamma$:
\begin{equation}
P\propto\rho^{\gamma}
\end{equation}
so that the total energy is
\begin{equation}
e=\frac{P}{\gamma-1}+\frac{1}{2}\rho u^{2}.
\end{equation}
%and the sound speed is given by
%\begin{equation}
%c_{\rm s} = \sqrt{\frac{\partial{P}}{\partial\rho}} = %\sqrt{\frac{\gamma{P}}{\rho}}.
%\end{equation}

\subsubsection{Hydrodynamic scheme}
\label{sec-dsa-hydro-scheme}

Our hydrodynamic scheme is adapted from the one used in \cite{Downes1998a}. It uses an Eulerian finite-volume approach, which allows us to take advantage of the conservative (hyperbolic) form of the Euler system (see eg \citealt{LeVeque1998a}). In this approach the functions $X(x)$ (where $X$ denotes any of the three components of $\vec{X}$ and $x$ is the space variable) are replaced by piece-wise approximations $X_i$ (where $i$ denotes the cell index). To update the $X$ values from time $n$ to time $n+1$ we readily discretize equation~(\ref{eq-cons}) as
\begin{equation}
\frac{V_i X_{i}^{n+1} - V_i X_{i}^{n}}{\delta{t}} = + S_{i-\frac{1}{2}}F_{i-\frac{1}{2}}^{n+\frac{1}{2}} - S_{i+\frac{1}{2}}F_{i+\frac{1}{2}}^{n+\frac{1}{2}}
\end{equation}
where $V_{i}$ is the volume  of cell $i$, $S_{i-1/2}$ ($S_{i+1/2}$) is the surface at the left (right) of cell $i$ and $F_{i-1/2}^{n+1/2}$ ($F_{i+1/2}^{n+1/2}$) is the flux of $X$ through this surface during the time-step $\delta{t}$. As $F_{i+1/2}^{n+1/2} = F_{(i+1)-1/2}^{n+1/2}$ this scheme has the interesting property that (but for boundary effects) the $X$ quantities are numerically conserved\footnote{
We note that the Godunov scheme works with the \emph{conservative} variables $\rho$, $\rho u$, $e$, whereas the Riemann solver works with the \emph{primitive} variables $\rho$, $u$, $P$. The pressure is computed as the difference between the total and kinetic energy of the fluid: $P = (\gamma-1)\left(e - 0.5\rho u^2 \right)$ so that in strong shocks where $0.5\rho u^2 \gg P$ and $e \sim 0.5\rho u^2$ its evaluation might become rather imprecise. Actually such a scheme can produce some fake "negative pressures" which have to be corrected to some fixed convenient $P_{min}$ (see eg \citealt{LeVeque1998a}). To prevent that problem \cite{Kang2002a} decided to add another equation specifically for the pressure, the equation of conservation of the "modified entropy" $S = P / \rho^{\gamma-1}$ (note that this equation is not valid at the shock, so that they need to combine it appropriately with the usual energy equation).
}, which is critical for computing the correct velocity of shock discontinuity.
Following \cite{Godunov1959a} the fluxes $F_{i\pm1/2}^{n+1/2}$ are computed using the values of $X$ at the interfaces computed by a Riemann solver (we use an exact Riemann solver with a linear approximation in smooth regions).

The code is made second order in time by using a leap-frog scheme, and is made second order in space by extrapolating the $X$ quantities at the cell interfaces before running the Riemann solver (doing slope reconstructions, which requires "slope limiters" to preserve monotonicity, see \citealt{Leer1977a}).

We recall that this scheme is subject to the usual Courant condition 
\begin{equation}
\label{eq-courant-adv-x}
\delta{t}_{\rm adv,x} < \frac{\delta{x}}{u_{\rm max}}
\end{equation}
where $u_{\rm max} = \left( \left|u\right| + c_{\rm s} \right)_{\rm max} $ is the maximum physical signal speed at the time considered.

\subsection{Particle Acceleration}
\label{sec-dsa-kino}

Particles that are energetic enough to be scattered off magnetic irregularities decouple from the thermal plasma (and are called CR). It is assumed that this scattering is strong enough that the CR are isotropised in both the upstream and downstream reference frame.
%, which allows them to catch up with the shock many times. 
We then describe CR through their isotropic distribution function $f(\vec{x},p,t)$ defined so that the CR density is
\begin{equation}
n(\vec{x},t)=\int_p{f(\vec{x},p,t)4\pi p^2 dp}
\end{equation}
and which obeys the following Fokker-Plank equation:
\begin{equation}
\label{eq-dce-f}\frac{\partial{f}}{\partial{t}} + \frac{\partial{uf}}{\partial{x}} = \frac{\partial}{\partial{x}}\left(D\frac{\partial{f}}{\partial{x}}\right) 
+ \frac{1}{3p^2} \frac{\partial{p^3f}}{\partial{p}} \frac{\partial{u}}{\partial{x}}.
\end{equation}
The second l.h.s. term represents convection : CR are advected in space as they are bound to the fluid by scattering off the magnetic waves present in it. Note that we neglect here the movements of the scattering centres (the waves) with respect to the fluid (we suppose $V_A \ll u$, where $V_A$ is the Alfven speed), so that we don't consider diffusion in momentum (known as Fermi~II acceleration). The first r.h.s. term represents space diffusion of the CR resulting from their scattering off the magnetic waves, conveniently described by a diffusion coefficient $D(x,p)$ (on which we shall focus in section~\ref{sec-scales}). The second r.h.s term represents adiabatic compression of the fluid, the velocity divergence $\frac{\partial{u}}{\partial{x}}$ being the engine of the particles acceleration.

\subsubsection{Kinetic scheme}
\label{sec-dsa-kino-scheme}

From now on we consider that CR are protons (of mass $m_p$), and we express CR momenta in $m_p c$ units (and CR velocities in $c$ units, CR energy and pressure in $m_p c^2$ units).

Following \cite{Falle1987a} we work with $g=p^4f$ and $y=\ln(p)$ for numerical convenience. We then rewrite equation~(\ref{eq-dce-f}) as
\begin{equation}
\label{eq-dce-g}
\frac{\partial{g}}{\partial{t}} + \frac{\partial{ug}}{\partial{x}} = \frac{\partial}{\partial{x}}\left(D\frac{\partial{g}}{\partial{x}}\right) 
- u_y \frac{\partial{g}}{\partial{y}} + u_y g
\end{equation}
where $u_y = -\frac{1}{3}\frac{\partial{u}}{\partial{x}}$ appears as a y-advection velocity (note also the new source term $u_y g$). Our kinetic scheme follows the one presented by \cite{Falle1987a}, but for the fact that we use here a Eulerian code so that we have to deal with space advection. In fact the particle transport is done in two steps using the operator splitting technique. Space transport (l.h.s of equation~(\ref{eq-dce-f})) is embedded into the hydrodynamic Godunov module: CR are transported with the fluid as a passive tracer during each hydrodynamic step. Then diffusive acceleration (r.h.s. of equation~(\ref{eq-dce-f})) is solved using a separate finite-difference module. The momentum variable $y=\ln(p)$ is linearly discretized with step $\delta{y}$, so that in each space cell (indexed by $i$) we have the full piece-wise spectrum of the particles (indexed by $j$). Ignoring the space transport terms, equation~(\ref{eq-dce-g}) is discretized as
\begin{eqnarray}
\label{eq-dce-g-num}
%\frac{g_{i,j}^{n+1}-g_{i,j}^{n}}{\delta{t}} & = & \omega^{n} \frac{D_{i+\frac{1}{2}}(g_{i+1,j}^{n}-g_{i,j}^{n}) - D_{i-\frac{1}{2}}(g_{i,j}^{n}-g_{i-1,j}^{n})}{\delta{x}^2} \nonumber \\
%& + & \omega^{n+1} \frac{D_{i+\frac{1}{2}}(g_{i+1,j}^{n+1}-g_{i,j}^{n+1}) - D_{i-\frac{1}{2}}(g_{i,j}^{n+1}-g_{i-1,j}^{n+1})}{\delta{x}^2} \nonumber \\
%& - &  {u_y}_{i}^{n} \frac{g_{i,j^+}^{n}-g_{i,j^-}^{n}}{\delta{y}} + {u_y}_{i}^{n} g_{i,j}^{n}
\frac{g_{i,j}^{n+1}-g_{i,j}^{n}}{\delta{t}} & = & \omega^{n} {\rm diff}_i^{n}  + \omega^{n+1} {\rm diff}_i^{n+1} \nonumber \\
& - &  {u_y}_{i}^{n} \frac{g_{i,j^+}^{n}-g_{i,j^-}^{n}}{\delta{y}} + {u_y}_{i}^{n} g_{i,j}^{n}
\end{eqnarray}
with
\begin{equation}
{\rm diff}_i^{n} = \frac{D_{i+\frac{1}{2}}(g_{i+1,j}^{n}-g_{i,j}^{n}) - D_{i-\frac{1}{2}}(g_{i,j}^{n}-g_{i-1,j}^{n})}{\delta{x}^2}
\end{equation}
where we have allowed for a space-dependent diffusion coefficient $D$ to be evaluated at the cells interfaces $i\pm1/2$, and where $\omega^n$ are coefficients which define the particular numerical scheme (see below).

We first comment on the last two r.h.s terms, representing CR energy gain. Y-advection is discretized using an upwind scheme, so that $j^+=j$ and $j^-=j-1$ when $u_y>0$. This leads to a Courant condition 
\begin{equation}
\label{eq-courant-adv-y}
\delta{t}_{{\rm adv},y} < \frac{\delta{y}}{{u_y}_{\rm max}}
\end{equation}
usually slightly more restrictive than the hydrodynamic condition~(\ref{eq-courant-adv-y}): in that case we just sub-cycle DSA according to the hydrodynamic time-step $\delta{t}_{\rm adv,x}$.

The Courant condition for the diffusive terms with a fully explicit schemes 
now reads
\begin{equation}
\label{eq-courant-diff-x}
\delta{t}_{{\rm diff},x} < \frac{\delta{x}^2}{2D_{\rm max}}
\end{equation}
which is much more restrictive than the advection condition, because of its quadratic dependence on the space resolution, and because $D$ is an increasing function of $p$ (see section~\ref{sec-scales-diff-coeff}).  Hence exploring acceleration to higher maximum momenta requires lowering $\delta{t}_{{\rm diff},x}$. To overcome this limitation we use  implicit schemes which are not limited by the Courant condition\footnote{
We have also investigated the Super-Time-Stepping (STS) method which allows explicit schemes to overcome the Courant condition (see eg \citealt{Alexiades1996a}). However, this didn't prove as time-saving as implicit schemes.
} (at the cost of more involved computations, and with the risk of loosing accuracy control). As seen from equation~(\ref{eq-dce-g-num}) our scheme can be explicit ($\omega^{n}=1, \omega^{n+1}=0$), implicit ($\omega^{n}=0, \omega^{n+1}=1$), or both, the special case $\omega^{n}=1/2, \omega^{n+1}=1/2$ being known as the Crank-Nicholson scheme. This scheme is of particular interest as it is the only one to be second order in both space and time. However it has some drawbacks too, as it may give unphysical negative values for small values of $\delta{t}$ (\citealt{Park1996a}). In that respect the fully implicit scheme is more robust, but it is less accurate.

Finally we need to define boundary conditions for space diffusion (for space advection CR share the fluid density space boundary conditions) : they can be either \emph{no flux} (reflecting boundaries) or \emph{no particle} (absorbent boundaries). Regarding momentum boundary conditions we simply impose that $g(p_{\rm min})=g(p_{\rm max})=0$.

\subsubsection{Injection process}
\label{sec-dsa-kino-injection}

We also need to address the problem of the injection of CR from the fluid. As said before the CR are initially nothing but the high energy particles of the thermal distribution. However the problem arising in our two populations approach of filling the gap between the fluid and the kinetic regimes of the particles is a rather delicate one.

Injection has usually been parametrized by two quantities (see eg \citealt{Falle1987a}, \citealt{Kang1991a}): the fraction, $\eta$, of the particles crossing the shock becoming CR, and the momentum $p_{\rm inj}$ at which they are injected. The latter can be fixed or defined by $p_{\rm inj} = \xi p_{\rm th,2}$ where $p_{\rm th,2} = \sqrt{2 m_p k_B T_2}$ is the mean downstream thermal momentum (or alternatively by $v_{\rm inj} = \xi' c_{\rm s,2}$ where $v$ is the particle speed and $c_{\rm s,2} = \sqrt{\gamma k_B T_2 / m_p}$ is the downstream sound speed). $\xi$ is expected to be in the range $2-4$ ($\xi'/\xi=\sqrt{2/\gamma} \simeq 1.1$ for $ \gamma=5/3$). The parameter $\eta$ is less constrained, with a typical order of magnitude of $\eta = 10^{-3}$.

However \cite{Malkov1998a} have developed a self-consistent analytical model of injection of suprathermal particles, known as the ``thermal leakage'' mechanism. \cite{Gieseler2000b} have implemented it in the CRASH code through the use of a ``transparency function'' which connects the thermal and supra-thermal distributions. They have then only one remaining free parameter (the ``wave amplitude''), which is rather well constrained -- at least for strong shocks for which the self-consistent injection rate often produces quickly and strongly modified shocks which are difficult to handle numerically (see \citealt{Kang2002a}). We adopt in our code a simplified recipe based on this thermal leakage mechanism, proposed by \cite{Blasi2005a}: given a value of $\xi$ we compute $\eta$ as
\begin{equation}
\label{eq-eta-blasi}
\eta_{\rm B}(\xi) = \frac{4}{3\sqrt{\pi}} (r-1) \xi^3 \exp(-\xi^2).
\end{equation}
We note that there is also only one remaining parameter, $\xi$, which is also well constrained -- but $\eta$ has a strong dependence on it. Although very simple to implement, this recipe allows a self-consistent description of the leakage of high energy thermal particles to the suprathermal population. Note that $\eta$ is a function of the shock compression ratio $r$ and thus a function of time in the case of a modified shock (see section~\ref{sec-dsa-kino-nonlinear}): the $r-1$ factor acts as an injection regulator.

We thus add a source term to the r.h.s of equation~(\ref{eq-dce-f}):
\begin{equation}
Q_{\rm inj}(x,p,t) = \eta\left(\xi(t)\right) \frac{\partial F(x_{\rm S},t)}{\partial x} G(x-x_{\rm S}) \delta(p-p_{\rm inj})
\end{equation}
where $x_{\rm S}$ is the shock location, $F(x_{\rm S}) = {\left(\rho{u}\right)}_{\rm S}/m_p$ is the particles flux through the shock (evaluated by the code in the shock frame), $G(x) = \frac{1}{\sqrt{2\pi}\epsilon} \exp\left(-\frac{x^2}{2\epsilon^2}\right)$ is a Gaussian distribution which spreads the injection around the shock location and $\delta$ is the Dirac distribution. Moreover we have to take account of the fact that these particles are extracted from the thermal population. As usual we neglect the inertia of the fresh CR, but we remove their energy from the fluid: we add a corresponding sink term to the fluid energy equation:
\begin{equation}
S(x,t) = \eta\left(\xi(t)\right) \frac{\partial F(x_{\rm S},t)}{\partial x} G(x-x_{\rm S}) \frac{m_p u_{\rm inj}^2}{2}.
\end{equation}

\subsubsection{Test-particle acceleration}
\label{sec-dsa-kino-linear}

In the linear test-particle regime the DSA mechanism is well known: theory provides various results we have used to validate our code and we summarize here.

Particles injected at constant momentum $p_{inj}$ build a power-law spectrum
\begin{equation}
\label{eq-pl}
f(p) = f_0 p^{-s}
\end{equation}
of slope
\begin{equation}
s = \frac{3r}{r-1}.
\end{equation}
where $r$ is the shock compression ratio (taken as 4 for strong shocks).
For a strong shock, then, $s=4$, for weaker shocks $s>4$ (that is, $f \propto p^{-4}$ is the hardest spectrum achieved by linear acceleration by a single shock -- without losses).
If the injection rate, $\eta$, is constant and the upstream medium is homogeneous then the spectrum normalisation is
\begin{equation}
f_0 = s \frac{\eta \rho_1}{4\pi m_p p_{\rm inj}^3}.
\label{eq-f0}
\end{equation}

This spectrum extends from $p_{\rm inj}$ to a maximum momentum $p_{\rm max}$ controlled by the scattering of the CR and thus in our model by the diffusion coefficient $D$. If we assume for the sake of simplicity that $D$ is constant in space and has a simple power-law dependence on $p$: $D(p)=D_0 p^{\alpha}$ then $p$ grows in time as
\begin{equation}
\label{eq-pmax}
\ln\left(\frac{p(t)}{p_{\rm inj}}\right) =
\left| 
\begin{array}{ll}
\displaystyle \frac{1}{\alpha} \ln\left(1+\alpha\frac{t}{t_{\rm acc}(p_{\rm inj})}\right) & \alpha\neq0\\
\displaystyle \frac{t}{t_{\rm acc}(p_{\rm inj})} & \alpha=0
\end{array}
\right.
\end{equation}
where
\begin{equation}
\label{eq-tacc0}
t_{\rm acc}(p) = \frac{3}{u_1-u_2} \left( \frac{1}{u_1} + \frac{1}{u_2} \right) D(p)
\end{equation}
is the characteristic acceleration time-scale at momentum $p$ (see \citealt{Drury1983a}).

\subsubsection{The non-linear problem: Particle back-reaction and shock modification}
\label{sec-dsa-kino-nonlinear}

It has been noted since the early developements of the DSA theory that the CR pressure, defined as
\begin{equation}
P_{\rm cr} = \int_p \frac{pv}{3} f(p) 4\pi p^2 dp = \frac{4\pi}{3} \int_y \frac{p^2}{\sqrt{1+p^2}} g(y) dy
\label{eq-pc}
\end{equation}
grows without limit in the linear regime, which implies that some backreaction process occurs. In fact as the CR diffuse upstream of the shock their pressure gradient induces a force which pre-accelerates the fluid (see eg \citealt{Kirk1994a} and \citealt{Berezhko1999a}): there is formation of a so-called ``precursor'', a smooth and spatially extended ramp in the hydrodynamical profiles upstream of the shock. The shock itself is thus progressively reduced to a so-called ``sub-shock'', whose compression ratio is $r_{sub}<4$ (however, the overall compression ratio $r_{tot}$ from far upstream to far downstream can now be $>4$). Particle acceleration is then fully coupled with the shock evolution: CR accelerated by the shock modify its hydrodynamical structure, modifying in turn the acceleration process itself and thus their spectrum (which is no longer a perfect power-law, but becomes concave in shape as particles of different momenta explore different scales upstream of the shock and thus feel different compression ratios).

Thus we must add the CR pressure contribution in the fluid momentum equation as well as in the fluid total energy equation (we still neglect the CR inertia, this will be discussed at the very end of the article): we add a source term 
\begin{equation}
\vec{Q}\left(\vec{X}\right) = \left( \begin{array}{c}
0 \\
-\vec{\nabla}P_{\rm cr} \\
-u \vec{\nabla}P_{\rm cr}
\end{array} \right)
\end{equation}
on the r.h.s. of equation~(\ref{eq-cons}).

Numerically $P_{\rm cr}$ is evaluated as
\begin{equation}
{P_{\rm cr}}_i^n = \frac{4\pi}{3} \delta{y} \sum_j \frac{p_{i,j}^2}{\sqrt{1+p_{i,j}^2}} g_{i,j}^n.
\end{equation}
and its space gradient is computed using centred differences.
The particle back-reaction is done at each hydrodynamical step. If the hydrodynamical and kinetic time-scales are too different we may miss the actual CR back-reation. To prevent the CR decoupling from the fluid we make sure that within each hydrodynamical time-step the relative variation of the fluid energy $\Delta e/e$ due to the CR back-reaction is never higher than $10\%$.

\subsection{Test 1: a strongly modified shock}
\label{sec-dsa-test}

To test our code we start with the second test case from the pioneers of the kinetic approach \cite{Falle1987a} (hereafter FG).

\subsubsection{Test Design}
\label{sec-dsa-test-design}

The shock wave is generated by a supersonic piston: a piston of constant Mach number $M_{\rm p} = u_{\rm p}/c_{\rm s,1}$ generates a shock of constant Mach number $M_{\rm S} = u_{\rm S}/c_{\rm s,1}$ given by (eg \citealt{Landau1959a})
\begin{equation}
\label{eq-ms}
M_{\rm S} = \frac{\gamma+1}{4}M_{\rm p} \sqrt{1 + \left(\frac{\gamma+1}{4}M_{\rm p}\right)^2}.
\end{equation}
%Equation~(\ref{eq-ms}) can be inverted as :
%\begin{equation}
%\label{eq-mp}
%M_{\rm p} = \left( 1 - \frac{(\gamma-1)M_{\rm S}^2+2}{(\gamma+1)M_{\rm S}^2} %\right) M_{\rm S}.
%\end{equation}
We work in the piston frame: the upstream medium is given a velocity of $-u_{\rm p}$, using a ``reflecting'' left boundary condition. The piston is then fixed at $x=0$ in the simulation box, and the shock emerges out of this left boundary. The right boundary condition is chosen to be gradient zero.
The upstream medium is initially of constant density $\rho_1^0=1$ and pressure $P_1^0=1$. The piston Mach number is $M_{\rm p}=15$, so that $M_{\rm S}=100.5$, $u_{\rm S}=4.52$ in the upstream rest frame, and $r=3.55$.

There are no CR upstream initially ($P_{\rm cr}^0=0$). Particles are injected at a constant rate $\eta=0.0225$ at a variable momentum defined by $\xi'=2$ (see section~\ref{sec-dsa-kino-injection}). The piston beta ($\beta=u_p/c$) is adjusted so that initially $p_{\rm inj}^0=10^{-1}$ as done in FG. The momentum grid extends from $\log(p_{\rm min})=-3$ to $\log(p_{\rm max})=+4$, with a resolution (not critical here) $\delta{y}=0.23$ (that is 10 bins per decade). The diffusion coefficient is a power-law with a weak momentum dependence: $D\left({p}\right)\propto{p}^{0.25}$. Its normalisation is adjusted so that the simulation unit time is the acceleration time-scale at injection $t_{\rm acc}(p_{\rm inj})$ (see equation~(\ref{eq-tacc0})) as implicitly done in FG.

The simulation is run to $t_{\rm end}=40$ as in FG to show convergence of the coupled fluid-CR system (in the linear case we then expect CR to be accelerated to $p_{\rm max}=3.2<4$). The space box size equals the distance travelled by the shock during that time (at constant velocity $u_{\rm S} = 1.77$ with respect to the piston located at $x=0$) plus 10 times the diffusion length of the highest energy CR (corresponding to $\Lambda=10$ as defined by equation (\ref{eq-lambda-max})) that is $x_{\rm max}=250$. The space resolution is set to $\delta{x}=2,9.10^{-2}$ to achieve numerical convergence (corresponding to $\lambda=0.050$ at $p_{\rm min}$ (and $\lambda=0.037$ at $p_{\rm inj}^0$) as defined by equation (\ref{eq-lambda-min})). The hydrodynamic Courant number is set to 0.8, the kinetic scheme being sub-cycled by another factor of 2.

\subsubsection{Physical results}
\label{sec-dsa-test-results}

\begin{figure}
\includegraphics*{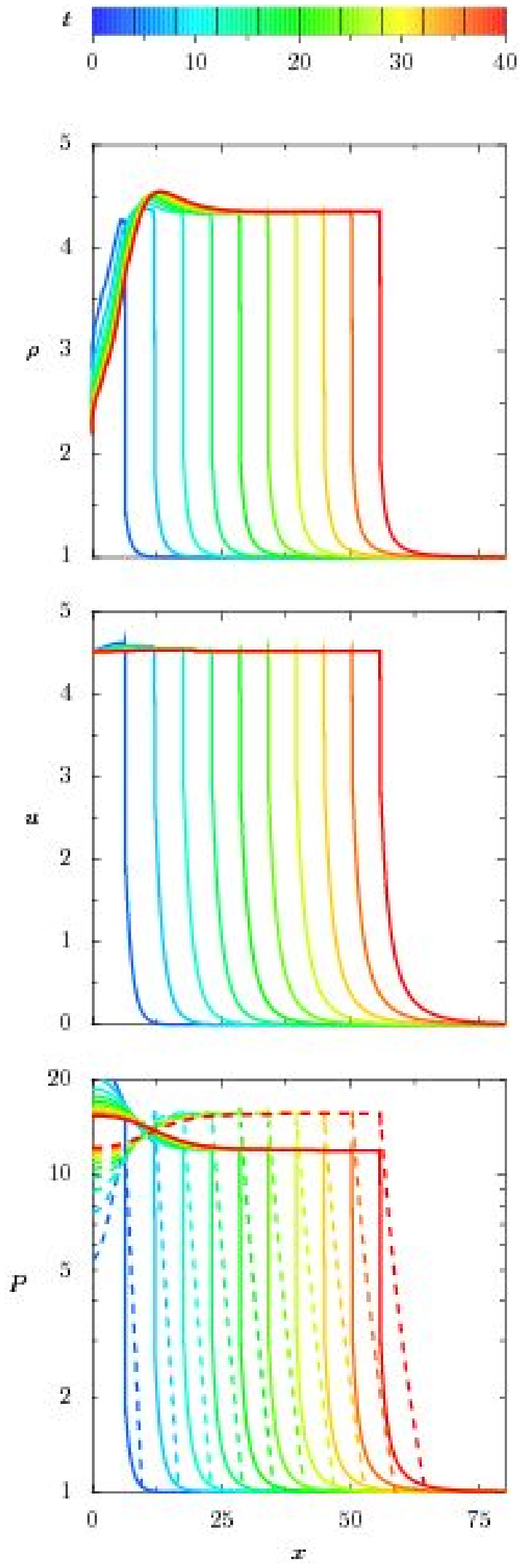}
\caption{Time evolution of the hydrodynamical profiles for a modified shock (with $D(p)\propto p^{0.25}$, see section~\ref{sec-dsa-test-design} for simulation details). Plotted are the fluid density $\rho$, velocity $u$ and pressure $P$ (the CR pressure $P_{\rm cr}$ is added dashed).}
\label{fig-test1-hydro}
\end{figure}

\begin{figure}
\includegraphics*{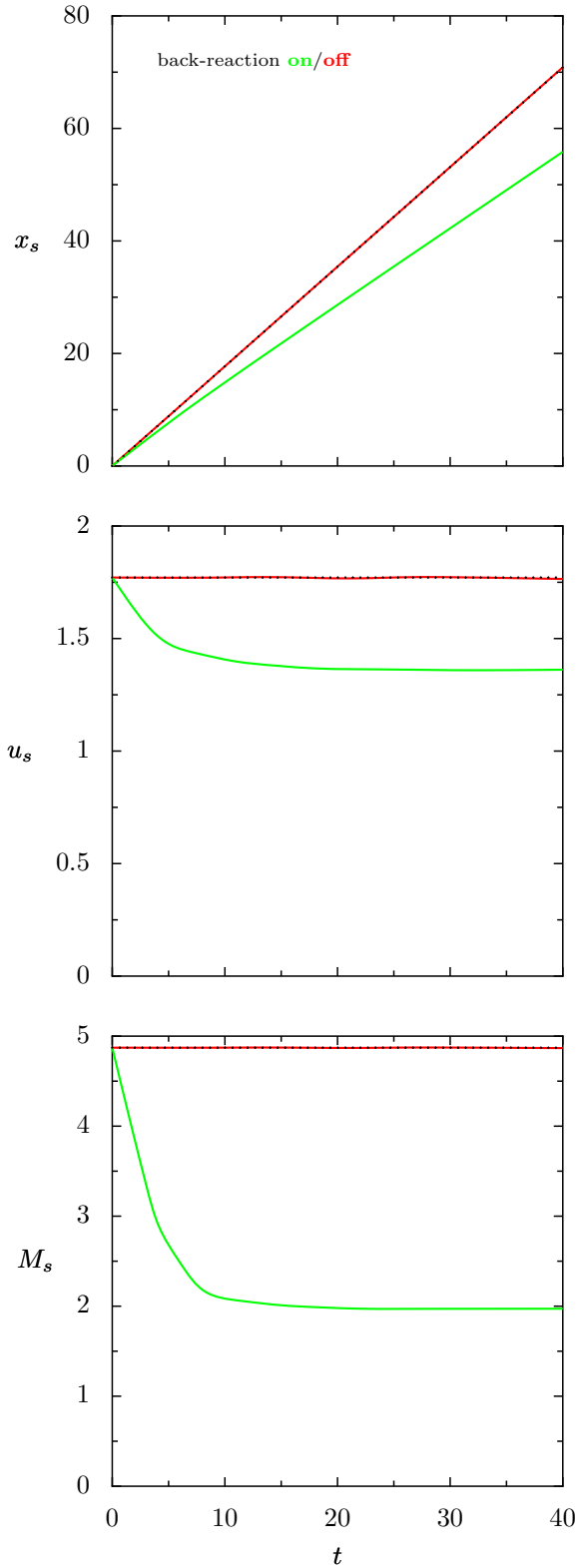}
\caption{Evolution of the strongly modified shock (in green) of figure~\ref{fig-test1-hydro}: the shock position $x_{\rm s}$, velocity $u_{\rm s}$ and Mach number $M_{\rm s}$ are plotted versus time. In red are added the results when CR back-reaction is turned off, which follow the theoretical evolution (dotted).}
\label{fig-test1-shock}
\end{figure}

\begin{figure}
\includegraphics*{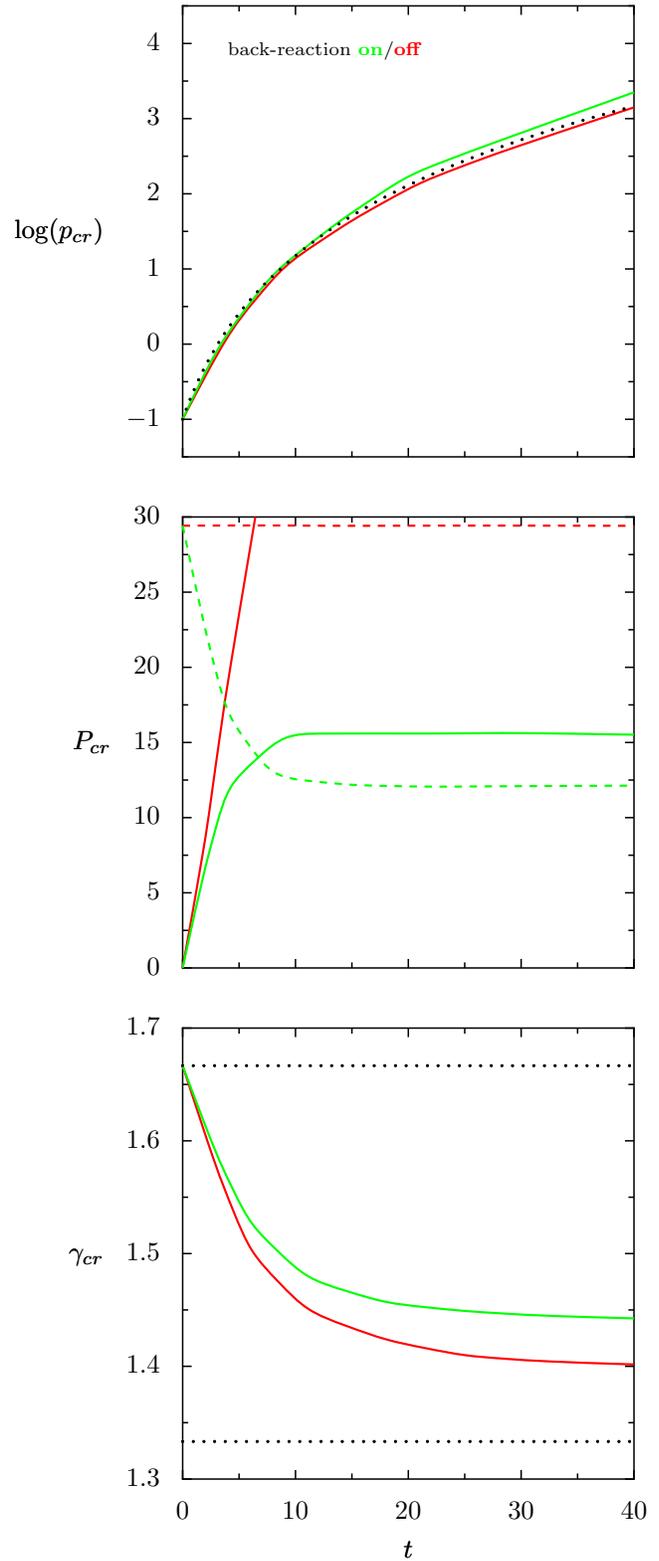}
\caption{Evolution of some key CR parameters for the modified shock (in green) of figure~\ref{fig-test1-hydro}: plotted versus time are the maximum momentum $p_{\rm cr}$ (its theoretical linear evolution is added dotted, see equation~(\ref{eq-pmax})), pressure $P_{\rm cr}$ (the fluid pressure is added dashed) and adiabatic index $\gamma_{\rm cr}$ (the non-relativistic ($5/3$) and ultra-relativistic ($4/3$) values are added dotted). In red are added the results when CR back-reaction is turned off.}
\label{fig-test1-kino-t}
\end{figure}

\begin{figure}
\includegraphics*{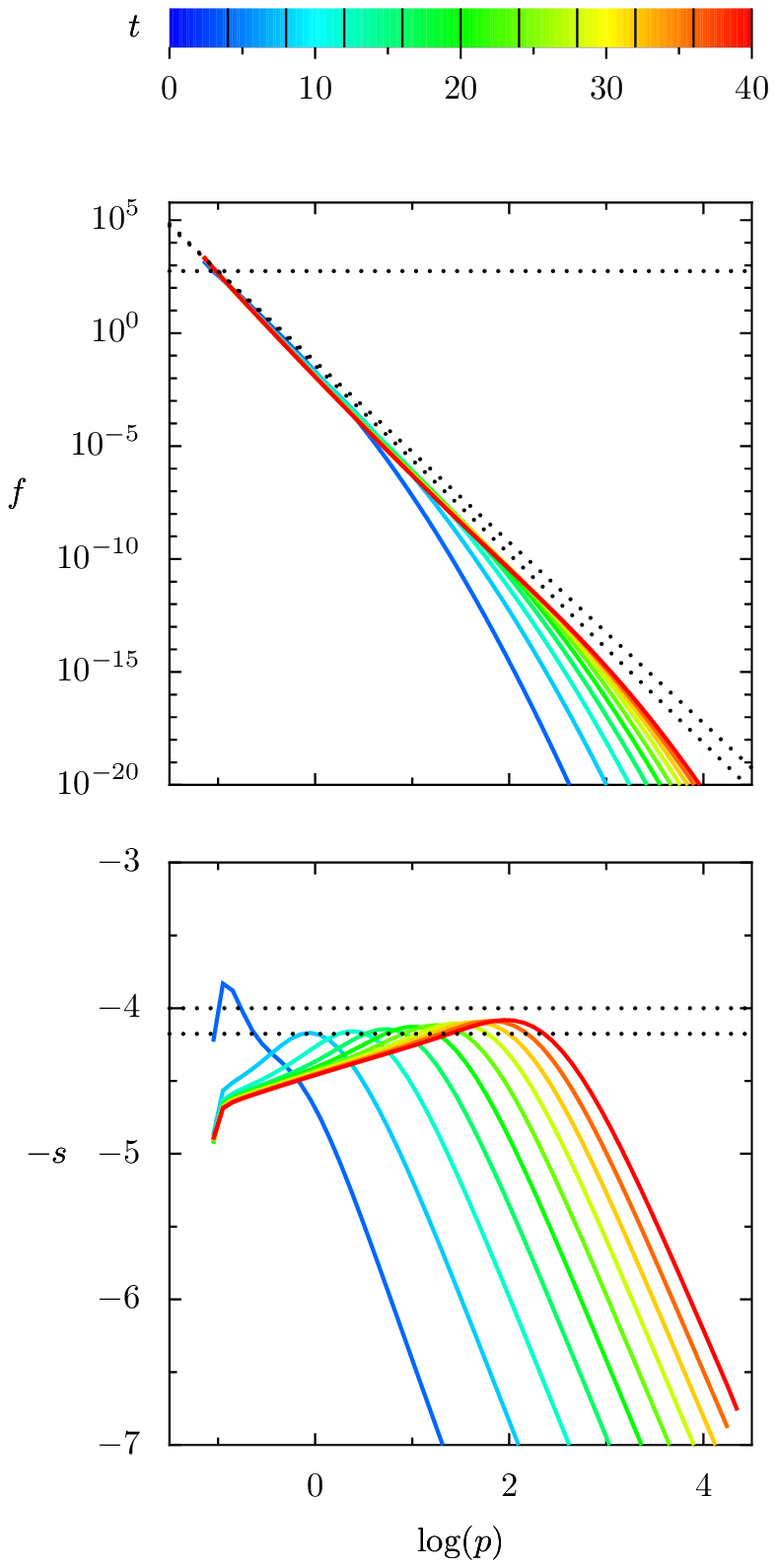}
\caption{Time evolution of the CR spectrum just downstream of the modified shock of figure~\ref{fig-test1-hydro}. The spectrum $f(p)$ is shown on top. On this plot the horizontal dotted line marks the theoretical spectrum normalization $f_0$ at injection (in the linear case) and the two other dotted lines are power-law spectra of slopes $s=4$ and $s_1=4.18=$ the theoretical linear slope (for this shock compression ratio $r=3.55$) and of same normalization $f_0$ at injection. These two remarkable slope values are also marked dotted on the bottom plot, which shows the spectrum logarithmic slope $-s=\partial\log(f) / \partial\log(p)$.}
\label{fig-test1-kino-p}
\end{figure}

The results of the simulation are presented in figures~\ref{fig-test1-hydro} to~\ref{fig-test1-kino-p}.

%\begin{figure}
%\includegraphics*{test1_hydro_shock.eps}
%\caption{Same as figure~\ref{fig-test1-hydro}, but in the shock frame instead of the upstream rest frame.}
%\label{fig-test1-hydro-shock}
%\end{figure}

Figure \ref{fig-test1-hydro} shows the hydrodynamical profiles for various times. The shock is visible as a discontinuity traveling to the right. As expected (see section~\ref{sec-dsa-kino-nonlinear}) it is smoothed by the presence of a precursor, visible in all profiles, caused by the presence of CR upstream of the shock -- their pressure, added dashed on the bottom plot, exceeds the fluid one. We see that after a quick initial adjustment the shock structure reaches a quasi-stationary state, as observed by FG.

Figure \ref{fig-test1-shock} shows the time evolution of the shock (in green). We see that after the quick initial adjustment the shock velocity is again almost constant, but slower than in the linear case (added in red).

Figure \ref{fig-test1-kino-t} shows the time evolution of the CR population (again with back-reaction switched on/off in green/red). In both cases the CR maximum momentum agrees well with the dotted theoretical linear prediction (the determination of $p_{\rm cr}$ is less reliable in the non-linear case, the differences seen are not conclusive). The CR pressure quickly converges thanks to the regulation effect of the back-reaction (as both the sub-shock mass flux and downstream temperature are reduced) whereas it grows linearly in time in the linear regime. The CR adiabatic index is computed as $\gamma_{\rm cr} = 1+P_{\rm cr}/E_{\rm cr}$ where $P_{\rm cr}$ is the "CR fluid" pressure given by equation~(\ref{eq-pc}) and $E_{\rm cr}$ is the "CR fluid" internal energy given by
\begin{equation}
E_{\rm cr} = \int_p K(p) f(p) 4\pi p^2 dp = 4\pi \int_y \frac{\sqrt{1+p^2}-1}{p} g(y) dy
\end{equation}
where $K(p)$ is the kinetic energy of a CR of momentum $p$. As expected $\gamma_{\rm cr}$ starts at the same value as the adiabatic index of a non-relativistic fluid $\gamma = 5/3$ (as CR are injected from the thermal pool) and goes down (constantly but more and more slowly) as CR are accelerated to high energies (tending to the adiabatic index $\gamma=4/3$ of a relativistic fluid).

The latter quantities are global (macroscopic) properties of the CR (seen as a fluid).
%But our code is a kinetic one, with a full CR spectrum in $p$ in each $x$ cell, from which $P_{\rm cr}$ and $\gamma_{\rm cr}$ are derived. 
Figure \ref{fig-test1-kino-p} shows the CR spectrum (and its slope) just downstream of the shock at some given times. We see how DSA progressively builds the CR distribution. Note that the time discretization is linear as in figure~\ref{fig-test1-hydro}, so we see that it takes more and more time to accelerate particles to higher energies: this is because of the diffusion coefficient dependence on $p$ (see equation~(\ref{eq-tacc0})). Initially the CR are injected at $\log(p_{\rm inj}^0)=-1$, with a normalization which agrees with the theoretical one in the linear case (see equation~(\ref{eq-f0})). But soon afterwards $p_{\rm inj}$ drifts slightly towards lower energies as the downstream temperature is reduced (see the bottom plot of figure~\ref{fig-test1-hydro}). The CR spectra extend over 20 orders of magnitude, they seem to approach a power-law form but the plot of their local slope clearly shows that they are actually concave: they are softer at low energies and harder at high energies than the theoretical linear slope $s_1=4.18$. This is another well-known feature of the acceleration by CR-modified shocks (see section~\ref{sec-dsa-kino-nonlinear}), due again to the energy dependence of the diffusion coefficient 
%, which implies that CR of different energies explore different regions around the shock and thus "see" different compression ratios (because of the precursor they create in the fluid, see figure~\ref{fig-test1-hydro}).

\subsubsection{Comparison with previous study}
\label{sec-dsa-test-comp}

The results presented here can be compared to the FG second test (section~4.2, especially their figure~7). Most notably the fluid and CR pressure have exactly the same time evolution, with a convergence at the same values within a few percents. The shock Mach number and CR adiabatic index also agree well.
This is to our knowledge the first direct comparison to the results of FG, and this success cross-validates the two codes. It proves that our code can handle well a strongly modified shock produced by the tight coupling of fluid hydrodynamics and CR diffusive acceleration.

\section{Diffusion scales and Adaptive Mesh Refinement}
\label{sec-scales}

We have already seen in the previous section how the energy dependence of the diffusion coefficient drives the main features of modified shocks. This is also the reason why realistic kinetic simulations of DSA are numerically a challenging problem: the difficulty is the potentially huge range of space and time scales which must be resolved. We now investigate this important issue in details, showing its physical grounds and its numerical answer.

\subsection{Diffusion}
\label{sec-scales-diff}

\subsubsection{The diffusion coefficient}
\label{sec-scales-diff-coeff}

In our model the scattering of CR off magnetic field fluctuations is represented by a diffusive term (first r.h.s term of equation~(\ref{eq-dce-f})), controlled by the diffusion coefficient $D$ which can be expressed as
\begin{equation}
D = \frac{1}{3} l_{\rm mfp} v
\end{equation}
where $v$ is the particle velocity and $l_{\rm mfp}$ its mean free path at this energy. A special case of interest is the so called "Bohm limit" (see eg \citealt{Kang1991a} and \citealt{Duffy1992a}) reached when $l_{\rm mfp} \sim r_{\rm g}$ where $r_{\rm g} = p / e B$ is the particle gyro-radius, that is when the particles are scattered within one gyro-period, meaning that the turbulence causing particle scattering is random on the scale $r_{\rm g}$. This constitutes a lower limit on the (parallel) diffusion coefficient, and on the acceleration time-scales. This Bohm limit has been widely favored in the literature. In that case $D \propto p v$ so that
\begin{equation}
D_{\rm B}(p) = D_0 \frac{p^2}{\sqrt{1+p^2}}
\end{equation}
where one can evaluate $\displaystyle D_0 = 3\times10^{22} cm^{2} s^{-1} \times \left(\frac{B}{1\mu G}\right)^{-1}$. The dependence of $D$ on $p$ is also frequently conveniently modelled by a simple power-law form
\begin{equation}
D_\alpha(p) = D_0 p^\alpha.
\end{equation}
We note that $D_B(p)$ reduces to such a power-law in the non-relativistic limit ($\alpha = 2$) and in the ultra-relativistic limit ($\alpha = 1$).

We can also consider a space-dependent diffusion coefficient. A common choice is
\begin{equation}
D(x,p) = \frac{\rho_1}{\rho(x)} D(p).
\end{equation}
This dependence on $\rho$ avoids the sound wave instability studied by \cite{Drury1986a} and mimics the compression of the magnetic field attached to the density field.

Note that otherwise the code doesn't handle explicitly the evolution in space and time of the magnetic field: some constant mean value of $B$ is assumed to get the normalization of $D$. Therefore we restrict here ourselves to the study of parallel shocks. We note that CR are thought to trigger themselves the field perturbations which make them diffuse, through various instabilities excited when they stream upstream of the shock. Thus it is possible to compute the diffusion coefficient $D(x,p,t)$ self-consistently from the CR distribution itself, adding the wave transports equations. We postpone this problem to a future work.

\subsubsection{The diffusion scales}
\label{sec-scales-diff-res}

Whatever the precise description used for $D$, the key point is that it is thought to be a growing function of $p$. As the CR momenta span many orders of magnitude (from say $p=10^{-2}$ to $p=10^6$ or even $p=10^9$ for galactic CR) this introduces a wide range of scales. The relevant time scale is the diffusive acceleration time-scale (see equation~(\ref{eq-tacc0})):
\begin{equation}
\label{eq-tacc}
t_{\rm acc}(p) \propto \frac{D(p)}{u_{\rm S}^2}.
\end{equation}
The relevant space scale is the diffusion length of the CR upstream of the shock:
\begin{equation}
\label{eq-xupst}
x_{\rm upst}(p) = \frac{D(p)}{u_{\rm S}}.
\end{equation}
The space scales range from the microscopic scale where the CR decouple from the fluid (of the order of a few thermal gyration lengths) to macroscopic scales (of the order of the supernova remnant radius for high energy CR, which then escape, thus limiting the $p_{\rm max}$ the remnant can achieve).

From a numerical perspective the resolution of the grid is then dictated by the diffusion of the lowest energy CR (we must ensure $\delta x \ll D(p_{\rm min})/u_{\rm S}$ to catch their dynamics well) whereas the size of the grid is dictated by the diffusion of the highest energy CR (we must ensure $x_{\rm max} \gg D(p_{\rm max})/u_{\rm S}$ not to lose them artificially). The ratio $D(p_{\rm max})/D(p_{\rm min})$ (and thus the number of cells $x_{\rm max} / \delta x$) may exceed ten orders of magnitude if $D(p) \propto p$, which is extremely demanding in terms of memory requirements and computing time. This is the reason why the first simulations were made with low $p$ dependence of $D$ ($\alpha = 0.25$ in \cite{Falle1987a} and in \cite{Kang1991a}), before exploring the Bohm regime (\citealt{Duffy1992a}) -- which was achieved by using more involved numerical techniques.

\subsection{Adaptive Mesh Refinement}
\label{sec-scales-amr}

\subsubsection{Principle}
\label{sec-scales-amr-principle}

Fortunately we need very high resolution ($\delta x$ small) only around the shock, as this resolution is required by the lowest energy particles only, which don't diffuse far away from the shock. More generally CR of a given energy require a certain space resolution on a certain space extended around the shock (the key parameter being $x_{\rm upst}(p)$). Hence the idea, pioneered by \cite{Duffy1992a} and developed by \cite{Kang2001a}, to implement techniques of Adapative Mesh Refinement (AMR) to allow the numerical resolution $\delta x$ to vary according to the needs of the CR that are likely to be found at a given location at a given time (see also \cite{Berezhko1994a} for a different approach). This allows correct handling of the transport of CR whereas considerably lowering the numerical requirements.

\subsubsection{Design}
\label{sec-scales-amr-design}

We adopt here the technique of nested grids (\citealt{Berger1984a}): $N$ sub-grids of increasing resolution are added to the base-grid around the shock\footnote{We haven't used tree-based AMR as this technique is much more complicated to implement and as its main advantage is its versatility but the situation we have to deal with is well defined.}. The resolution of the grid at level $k$ (base grid being level $0$) is $\delta r_k = \delta r_0 / R^i$ where $R$ is the refinement factor, taken as usual to be $R=2$.

The grid hierarchy is automatically designed by the code according to the CR diffusion properties as follows. The resolution at the last sub-level $N$ is adapted to the lowest energy CR, of momentum $p_{\rm min}=p_N$:
\begin{equation}
\left(\delta x\right)_N = \lambda \times x_{\rm upst}(p_N)
\label{eq-lambda-min}
\end{equation}
where $\lambda \ll 1$ (\cite{Kang1991a} suggest $\lambda = 0.05$, and this is indeed what test~1 of section \ref{sec-dsa-test} required for full convergence). The resolution of the $N-1$ sub-level will necessarily be $\delta x_{N-1} = R \times \delta x_N$. This resolution will be good enough for all particles of momentum above some $p_{N-1}$ so that $\delta x_{N-1} = \lambda \times x_{\rm upst}(p_{N-1})$. Then the sub-grid $N$ should take care of all CR of momenta from $p_{N}=p_{\rm min}$ to $p_{N-1}$, in particular it must contain them as they diffuse from the shock, so we set its half-size $\Delta x$ as
\begin{equation}
\left(\Delta x\right)_N = \Lambda \times x_{\rm upst}(p_{N-1})
\label{eq-lambda-max}
\end{equation}
with $\Lambda > 1$. The sub-grid $N$ is now fully defined by its size $\Delta x$ and resolution $\delta x$, the process is then iterated to the level $N-1$, the resolution of which is imposed by the resolution at level $N$ and the size of which is imposed by the resolution at level $N-2$, and so on. The total number of sub-grids is adjusted semi-empirically to maximize the AMR efficiency.

From now on the grids design will be conveniently described by the two parameters $\lambda$ and $\Lambda$ defined by relations~(\ref{eq-lambda-min}) and~(\ref{eq-lambda-max}). We use the same $\Lambda$ for all sub-grids, and we extend its definition to a base grid where $\Delta x$ now refers to the distance $x_{\rm max} - u_{\rm S} \times t_{\rm end}$ between the position of the shock at the end of the simulation and the position of the right physical boundary (so that in all cases $\Delta x$ is the minimum distance upstream of the shock).

\subsubsection{Algorithm}
\label{sec-scales-amr-algo}

The nested-grids algorithm is a recursive one: at each level (from top to bottom) we update the quantities on the whole grid, we run the same process at the sub-level, and we replace the grid coarse quantities by the sub-grid finer quantities (we also correct the coarse fluxes at the sub-grid interfaces to preserve the scheme conservation properties, see \citealt{Berger1998a}). We recall that the refinement is both a refinement in space (the resolution is divided by $R$) and a refinement in time (because of the Courant condition~(\ref{eq-courant-adv-x})). Note that the two operators in the code (the hydrodynamic one and the kinetic one) still operate conjointly at each level, as refining them separately would artificially decouple their effects. Note also that injection is done at bottom level only and propagates to all the upper levels as they are updated.

To set up the refinement the child grid must be given appropriate boundary conditions to match its parent profiles. Regarding the hydrodynamics these nested boundary conditions consist of a simple filling of the child's ghost with the corresponding values of its parent. Regarding the kinetic part we note that diffusion is controlled by interface fluxes, so that we enforce matching of the diffusion flux between the child and parent level. Note that this mechanism works whatever the diffusion coefficient scheme (be it explicit or implicit), but that the Crank-Nicholson scheme gets more sensitive because of such nested boundary conditions.

The grid hierarchy is set up around the shock position at start-up and moves with it over time. For the shock tracking to remain efficient with any number of grids levels we allow each sub-grid to move independently both in space and time. However a sub-grid can move only by $R=2$ of its cells to keep the simple 1 to 2 correspondence of the AMR refinement scheme (and only at the end of a complete refinement step).

\subsection{Test 2: adding Bohm scaling}
\label{sec-scales-test}

Here we extend test~1 using a more realistic (and demanding) diffusion model.

%\begin{figure}
%\includegraphics*{test2_hydro.eps}
%\caption{Time evolution of the hydrodynamical profiles for a modified shock (with $D(p)\propto p$, see sections~\ref{sec-scales-test-design} for simulation details). Plotted are the fluid density $\rho$, velocity $u$ and pressure $P$ (the CR pressure $P_{\rm cr}$ is added dashed).}
%\label{fig-test2-hydro}
%\end{figure}

%\begin{figure}
%\includegraphics*{test2_hydro_grids.eps}
%\caption{Same as figure~\ref{fig-test2-hydro}, with the AMR grids hierarchy over-plotted at each output time: dotted lines mark the center of the grids (which follow the shock), dashed lines mark the boundaries of the 3 nested sub-grids.}
%\label{fig-test2-hydro-grids}
%\end{figure}

\begin{figure}
\includegraphics*{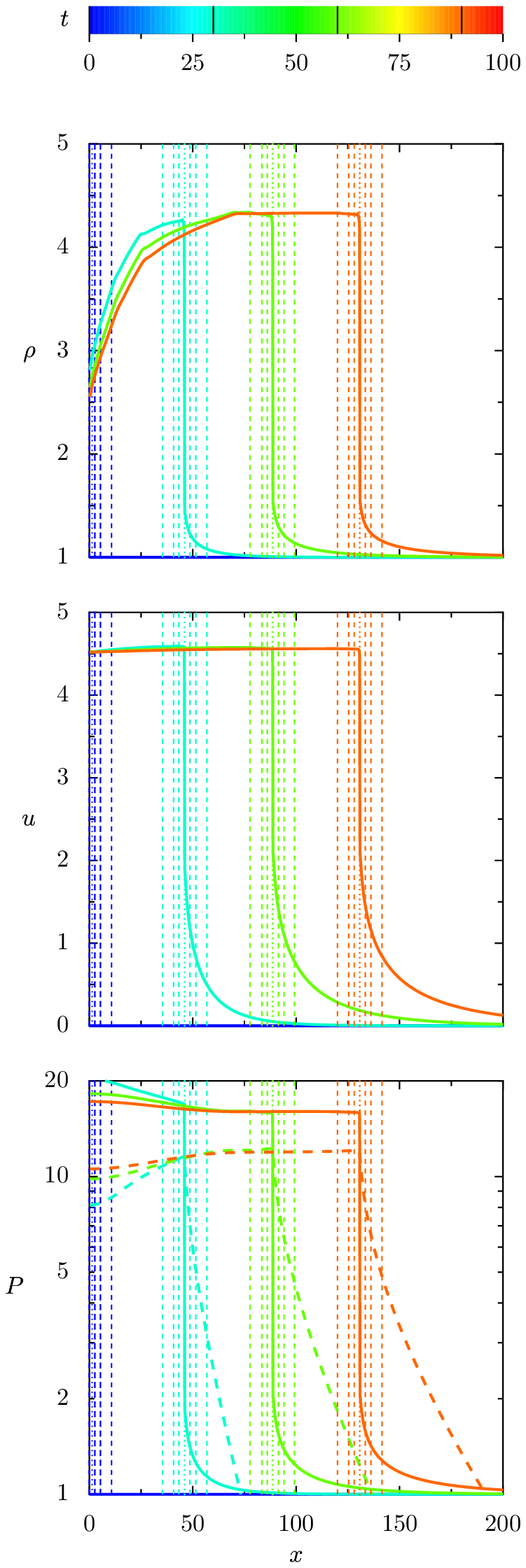}
\caption{Time evolution of the hydrodynamical profiles for a modified shock (with $D(p)\propto p$, see sections~\ref{sec-scales-test-design} for simulation details). Plotted are the fluid density $\rho$, velocity $u$ and pressure $P$ (the CR pressure $P_{\rm cr}$ is added dashed). The AMR grids hierarchy is over-plotted at each output time: dotted lines mark the center of the grids (which follow the shock), dashed lines mark the boundaries of the 3 nested sub-grids.}
\label{fig-test2-hydro-grids}
\end{figure}

\begin{figure}
\includegraphics*{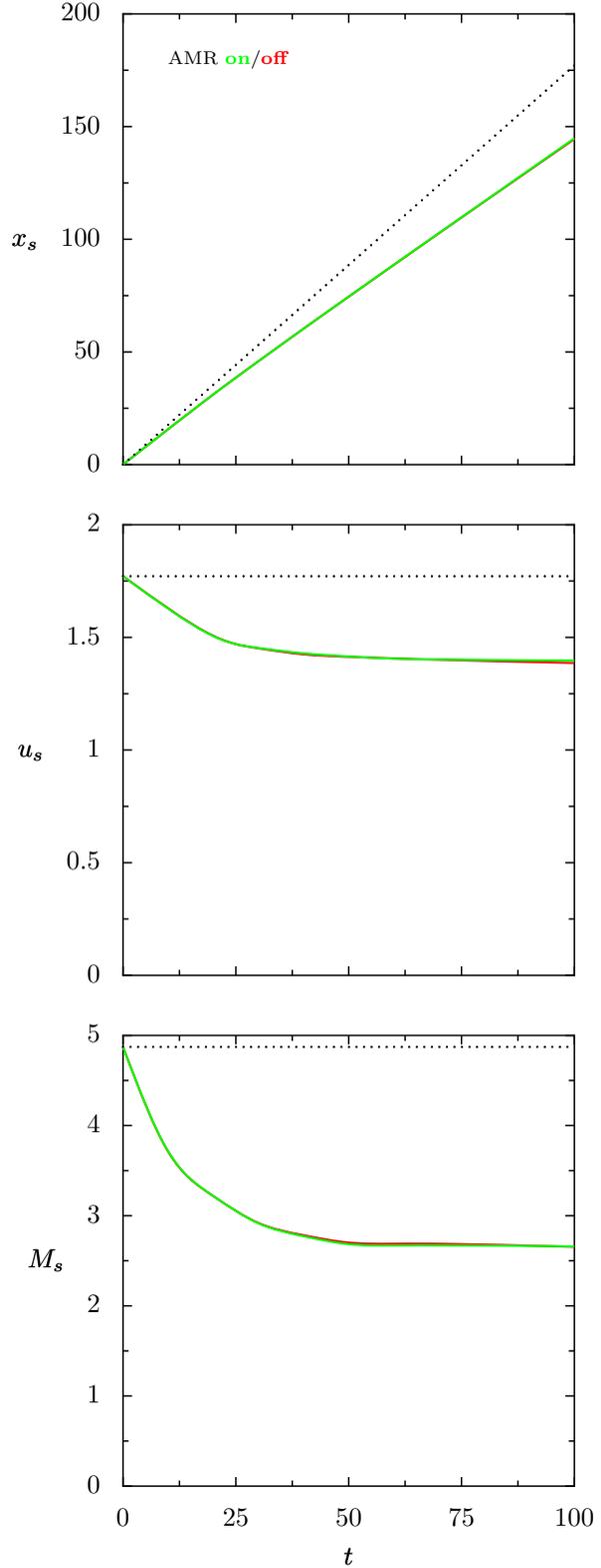}
\caption{Evolution of the strongly modified shock of figure~\ref{fig-test2-hydro-grids}: the shock position $x_{\rm s}$, velocity $u_{\rm s}$ and Mach number $M_{\rm s}$ are plotted versus time. The theoretical evolution in the non-modified case is added as the dotted line. Results obtained with and without using AMR are shown in green and red respectively.}
\label{fig-test2-shock}
\end{figure}

\begin{figure}
\includegraphics*{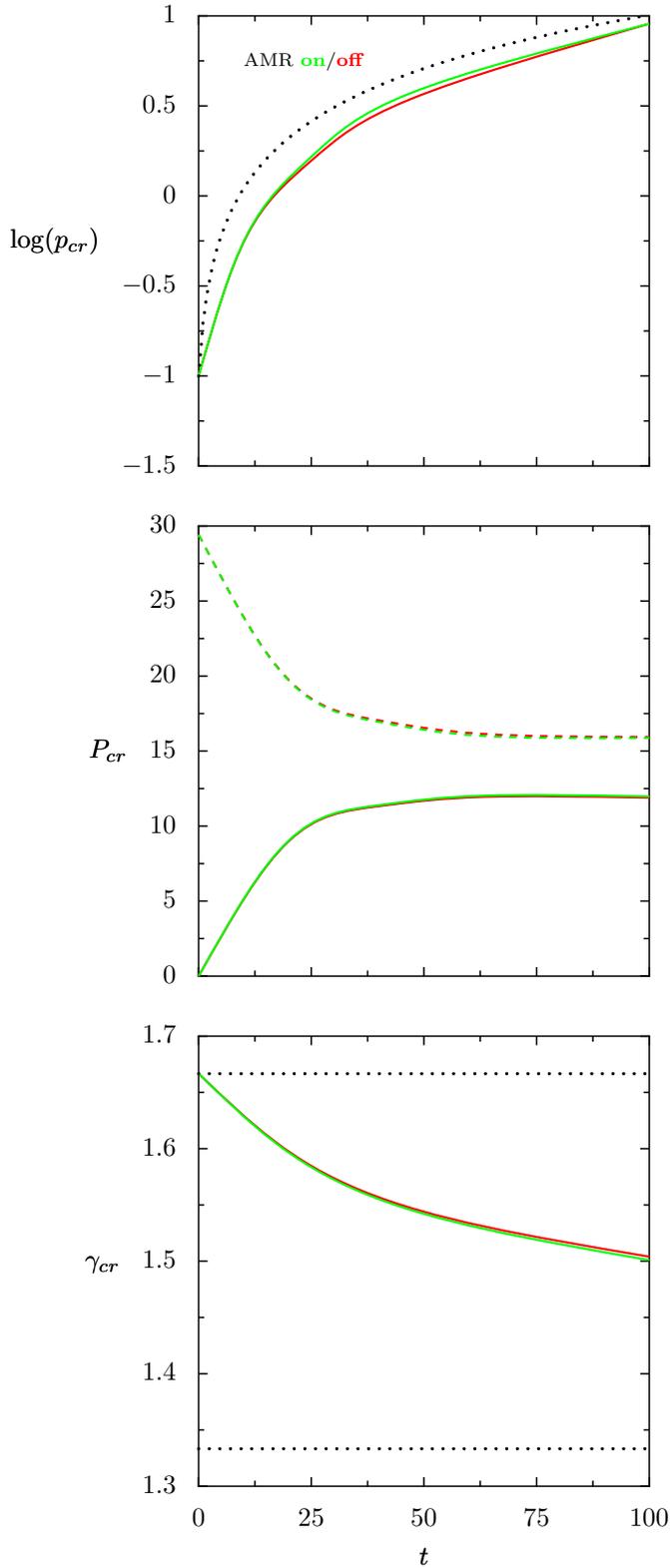}
\caption{Evolution of some key CR parameters for the modified shock of figure~\ref{fig-test2-hydro-grids}: plotted versus time are the maximum momentum $p_{\rm cr}$ (its theoretical linear evolution is added dotted, see equation~(\ref{eq-pmax})), pressure $P_{\rm cr}$ (the fluid pressure is added dashed) and adiabatic index $\gamma_{\rm cr}$ (the non-relativistic ($5/3$) and ultra-relativistic ($4/3$) values are added dotted). Results obtained with and without using AMR are shown in green and red respectively.}
\label{fig-test2-kino-t}
\end{figure}

\begin{figure}
\includegraphics*{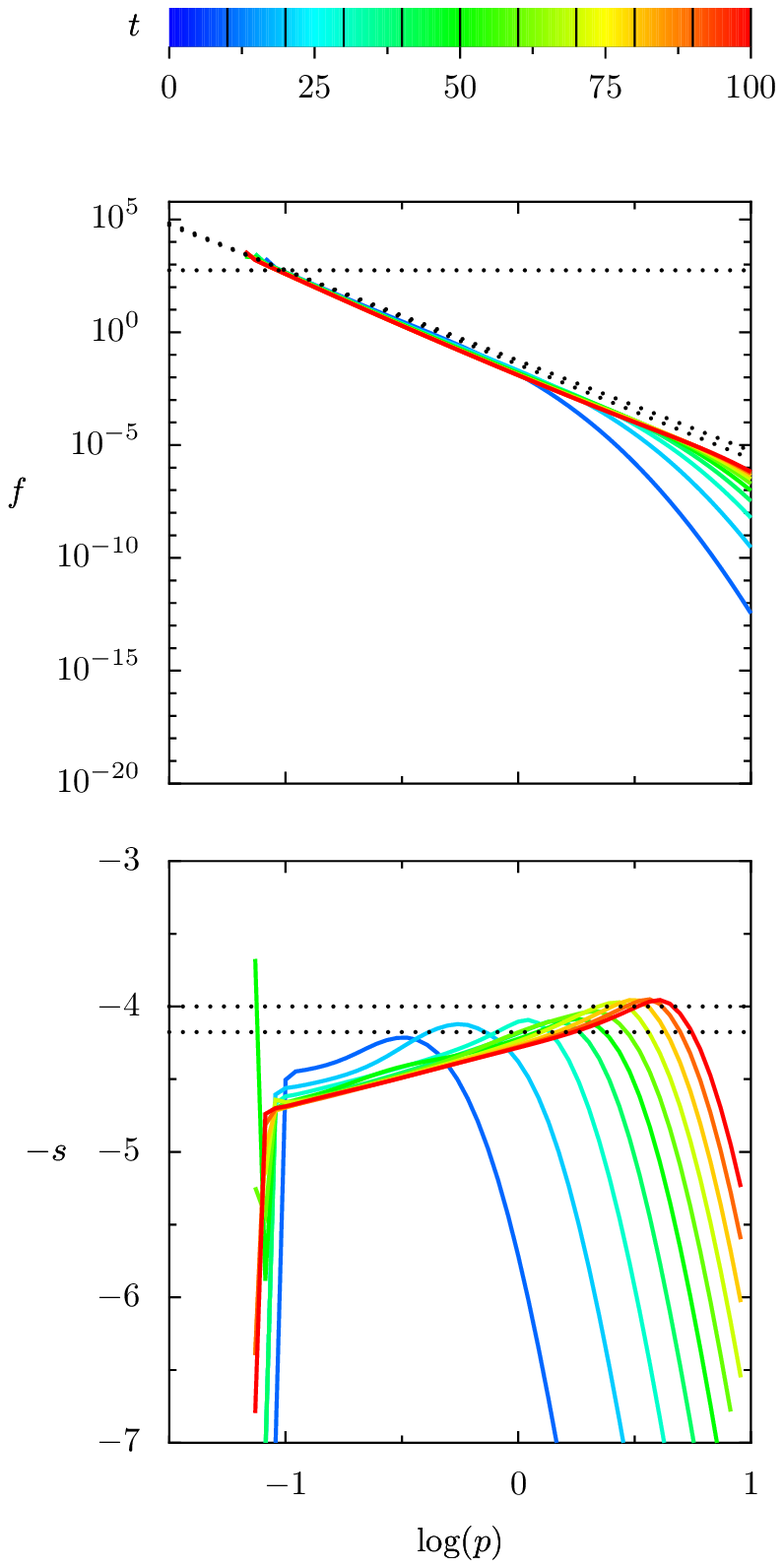}
\caption{Time evolution of the CR spectrum just downstream of the modified shock of figure~\ref{fig-test2-hydro-grids}. The spectrum $f(p)$ is shown on top. On this plot the horizontal dotted line marks the theoretical spectrum normalization $f_0$ at injection (in the linear case) and the two other dotted lines are power-law spectra of slopes $s=4$ and $s_1=4.18=$ the theoretical linear slope (for this shock compression ratio $r=3.55$) and of same normalization $f_0$ at injection.
These two remarkable slope values are also marked dotted on the bottom plot, which shows the spectrum logarithmic slope $-s=\partial\log(f) / \partial\log(p)$.}
\label{fig-test2-kino-p}
\end{figure}

\begin{figure}
\includegraphics*{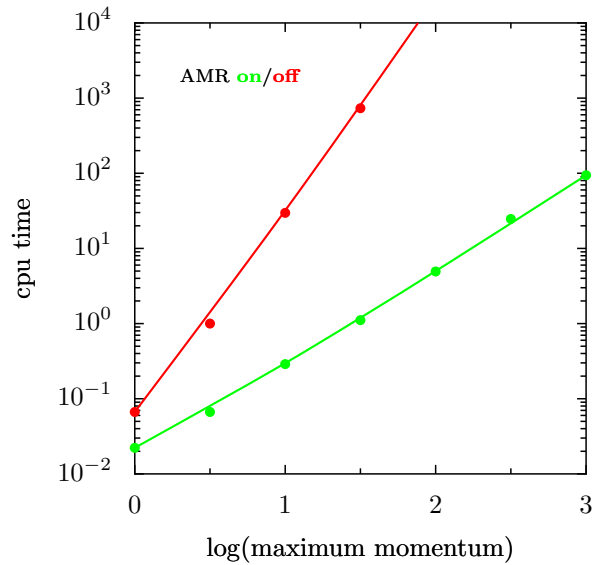}
\caption{Computing time as a function of the maximum momentum $p_{\rm max}$ (in logarithmic scale) for a Bohm-like diffusion ($D(p)\propto p$). The time unit on this plot is arbitrary, on an Itanium II processor the simulation to $\log(p_{\rm max})=1$ lasted 13.4~H with AMR off and 8~mn with AMR on. The CR are injected at $\log(p_{\rm inj})=-1$ at $t=0$, the minimum momentum is set to $\log(p_{\rm min})=-1.5$. The two sets of measures have been done with (green) and without (red) activating the automatic AMR scheme (see sections \ref{sec-scales-amr} and \ref{sec-scales-test-design} for details). Physical results for $\log(p_{\rm max})=1$ have been shown on figures \ref{fig-test2-hydro-grids} to \ref{fig-test2-kino-p}, with comparisons of results with and without AMR on figures \ref{fig-test2-shock} and \ref{fig-test2-kino-t}.}
\label{fig-test2-bench-amr}
\end{figure}

\subsubsection{Test Design}
\label{sec-scales-test-design}

The physical parameters are the same as in~\ref{sec-dsa-test-design}, but for the fact that we now use the relativistic\footnote{Using the exact Bohm scaling, that is $D\left({p}\right)\propto{p^2}$ at injection energies, considerably increases the numerical cost of the simulation without changing much the physical results, see eg \cite{Kang2006a}.} Bohm scaling for the diffusion coefficient, that is $D\left({p}\right)\propto{p}$ (see section~\ref{sec-scales-diff-coeff}). Numerically we then enter a new world, because of the requirements induced by equations~(\ref{eq-tacc}) (longer run) and~(\ref{eq-xupst}) (higher resolution). Simulation of test~1 runs within 2~hours at high resolution on a desktop-class processor. Now with the same $p_{\rm max}$ it wouldn't be even possible to allocate the grid in memory. We thus apply the AMR technique to lower the numerical requirements. We use here $\lambda = 0.3$ at $\log(p_{\rm min})=-1.5$ (that is $\lambda = 0.1$ at $\log(p_{\rm inj}^0)=-1$); and $\Lambda = 6$ for each sub-grid (as we have observed that for $\Lambda \geq 6$ the sub-grids nested boundary conditions for diffusion are indifferent for all the CR up to the momentum a sub-grid has to deal with) and $\Lambda = 10$ for the base grid. We use a better momentum resolution than in test~1: $\delta_y=0.1$ (that is 23 bins per decade). We've run different simulations with different maximum momentum $\log(p_{\rm max})$ ranging from 0 to 3 by steps of 0.5. We've run each simulation up to the time required for CR to reach this maximum momentum (derived from equation~(\ref{eq-pmax})), ranging accordingly from 10 to 10000.
% (the unit time is as for test~1 the acceleration time-scale at injection $t_{\rm acc}(p_{\rm inj})$ given by equation~(\ref{eq-tacc0})). 
We set the Courant number to 0.5 and we sub-cycle the DSA scheme a few times (all the more so since we are at a deep level). The number $N$ of sub-grids automatically dumped by the code ranges from 1 to 7 as $\log(p_{\rm max})$ rises from 0 to 3.

\subsubsection{Physical results}
\label{sec-scales-test-results}

Figures~\ref{fig-test2-hydro-grids} to~\ref{fig-test2-kino-p} show the results in the case $\log(p_{\rm max})=1$, $t_{\rm end}=100$, $N=3$ (note that on figure~\ref{fig-test2-hydro-grids} one sees how the grid hierarchy follows the shock). The global picture remains the same as in test~1: quick convergence to a quasi-steady state where the fluid and CR pressures are of the same order (here the CR pressure doesn't reach the fluid one). The effects of the dependence of $D$ on $p$, which were already visible in test~1, are now enhanced: the shock precursor is more extended (figure~\ref{fig-test2-hydro-grids}), the CR are more slowly accelerated towards high energies (figure~\ref{fig-test2-kino-p}).
%, where we recall that the time discretization is linear).

\subsubsection{AMR efficiency}
\label{sec-scales-test-bench}

On figures~\ref{fig-test2-shock} and~\ref{fig-test2-kino-t} the red curves show the results (in the case $\log(p_{\rm max})=1$) without activating AMR, that is with a single grid having the same size as the base grid (given by $\Lambda = 10$) and the same resolution as the deepest grid (given by $\lambda = 0.3$). The results can hardly be distinguished, which proves that AMR doesn't compromise the physical accuracy. And we show now that on the other hand it considerably lowers the numerical cost. Simulations described previously have been made (up to $p_{\rm max}=1.5$) without using AMR too. The computing time of each simulation with (green) and without (red) AMR is shown on figure~\ref{fig-test2-bench-amr}. In the case $\log(p_{\rm max})=1$ presented in section~\ref{sec-scales-test-results} the speed-up is of roughly 250; and it grows steadily as $p_{\rm max}$ grows.
% -- up to $\log(p_{\rm max})=2$ above which the code would simply be of no use without AMR in this configuration. 
We note that both curves are power laws, with an index $240\%$ times lower when AMR is activated. Thus the AMR technique is both very efficient and absolutely mandatory to address such difficult (realistic) problems.

\section{Multiple shocks and Parallelization}
\label{sec-multi}

Even using the numerical trick of AMR, the computing time of realistic DSA simulations can still be very high, especially if one wants a precise estimate of the CR spectrum over a wide range of momenta. This quickly becomes an unacceptable limitation if one wants to investigate the effects of multiple, successive shocks. In this section we present an original attempt to increase the code computing efficiency, aimed at increasing the number of simulated shocks in a given reasonable simulation time.

\subsection{Multiple shocks acceleration}
\label{sec-multi-msa}

In many astrophysical contexts CR are likely to experience many successive shocks: in chaotic stellar winds (\citealt{White1985a}), in rotating accreting flows (\citealt{Spruit1988a}), in radio sources with multiple hot spots (\citealt{Pope1996a}), in the galactic center (\citealt{Melrose1997a}), in the OB associations inside superbubbles (\citealt{Klepach2000a}, \citealt{Parizot2004a}), in the early cosmological flows (\citealt{Kang2005a}) and in galaxy clusters (\citealt{Brunetti2007a}). Many efforts have been made to better understand multiple diffusive shock acceleration (hereafter mDSA), but on quite particular cases, and clearly not to the same extent as single diffusive shock acceleration (hereafter sDSA). 

\subsubsection{Previous studies}
\label{sec-multi-msa-previous}

From a theoretical point of view, mDSA is well understood in the linear regime, as we can simply add the effect of a single shock. Being of astrophysical interest it has been investigated analytically since the early developments of the DSA theory (see eg \citealt{Eichler1980a}, \citealt{Blandford1980a}). The main result of mDSA is that the CR spectrum flattens progressively to a universal asymptotic power-law of index $s=3$ (regardless of the shock compression ratios, see eg \citealt{Pope1994a}). More involved analytical models have been developed including various effects such as second order Fermi acceleration, radiation losses (for electrons) and escape (\citealt{Schlickeiser1984a}, \citealt{Achterberg1990a}, \citealt{Schneider1993a}). But while there is a well-defined analytic framework for the linear mDSA, there is no such thing in the non-linear regime. However \cite{Bykov2001b} has developed a non-linear model of the acceleration by chaotic large scales fluid motions inside superbubbles, and \cite{Blasi2004a} has proposed a simple semi-analytical model including "seed" particles.

From a numerical point of view, as we have already seen the fully non-linear regime is quite well studied now, but in the single shock model: although feasible, multiple DSA has received extremely reduced attention so far. \cite{Gieseler2000a} have studied acceleration at multiple oblique shocks, but in the test-particle regime (they found the CR spectrum to harden substantially). Recently \cite{Kang2005a} have investigated the effect of an upstream CR pressure in the non-linear regime, but for single shocks (they concluded that it doesn't affect strong shocks much but may enhance the efficiency of weak ones).

Our aim with the code presented in this paper is to be able to study CR acceleration by multiple shocks in detail, in the full time-dependent non-linear regime.

\subsubsection{Inter-shock physics}
\label{sec-multi-msa-inter}

A new important point to consider when simulating multiple shocks is the fate of the CR \emph{between} two successive shocks. 

First the shocked fluid will decompress to recover its initial state, and the CR being bound to it through scattering will experience adiabatic decompression (see \citealt{Melrose1993a}): when the shocked fluid density is decreased by a factor $r$ the CR momenta are decreased by a factor $r^{1/3}$. We would like to insist on the physical importance of this decompression, whose inclusion is essential in the correct treatment of mDSA. For instance a sequence of identical shocks of ratio $r$ will produce in the linear regime a CR spectrum with the well-known $s=3$ slope if and only if the CR are decompressed by the corresponding $r^{1/3}$ factor between each shock: if we don't decompress them enough they will pile-up from the injection momentum, producing much harder spectra, and if we decompress them too much they will still form a power-law but a steeper one of slope $s>3$.

Apart from these energy losses due to decompression the CR might simply escape the system before the next shock occurs. In fact CR can be quite well confined in a medium thanks to their diffusion, whose typical length scale grows with the CR energy, so that the CR spectrum will be depleted from its highest part. In a given physical situation, given the typical time between two shocks and the diffusion experienced by CR during that time (not necessarily of Bohm type) we can estimate the maximum momentum of the remaining CR when the next shock arrives.

We also note that the second order Fermi acceleration mechanism (diffusion in momentum)
%, neglected in this paper regarding the interaction of CR with the shock,
 might become an important process between the shocks
% (particularly at low energies). 
However we won't investigate this possibility any further in this paper.

%Finally we want to stress that we simply have to change the details of the inter-shock physics in order to treat various astrophysical situations.

\subsubsection{Numerical treatment}
\label{sec-multi-msa-num}

The code is fully automated to run a sequence of shocks. Assuming that the fluid has enough time to decompress between two shocks, the same initial hydrodynamical conditions are used for each shock. The shock Mach number (and thus velocity and compression ratio) can however vary from shock to shock. At the end of each shock we take the downstream CR spectrum, modify it to take into account the inter-shock physics of section~\ref{sec-multi-msa-inter}, and pre-inject it in each space cell before launching the next shock. Although this mechanism seems simple we have to elucidate a few points.

First the very idea of a "downstream spectrum" supposes that we have reached some converged state downstream of the shock. According to \cite{Kang2005a} such a quasi-stationnary state is obtained once the CR are accelerated up to the relativistic regime ($p > m_p c$), which agrees with our own observations. In the following we consider that the time each shock is run is long enough so that each single shock can fully relax regarding particle acceleration (at least up to the maximum momentum we consider), so that the downstream CR pressure is well defined.

To mimic adiabatic decompression the CR spectrum $\ln(f)$ is shifted in $y=\ln(p)$ by
\begin{equation}
\Delta{y}(r) = \frac{1}{3}\ln(r)
\end{equation}
towards lower energies. Lost values of $f$ below $y_{\rm min}$ are simply discarded. We have checked by lowering the value of $y_{\rm min}$ that they don't influence the overall subsequent spectrum evolution. Missing values of $f$ between $y_{\rm max} - \Delta{y}$ and $y_{\rm max}$ are filled by linearly extrapolating the slope from the end of $f$. This treatment of the gap gives the best accuracy at high energies in the linear regime. As emphasised in section~\ref{sec-multi-msa-inter} to obtain correct results we need to resolve precisely this decompression shift and thus to use a high resolution in momentum: $\delta{y} \ll \Delta{y}$. We have found that in order to obtain exactly $s=3$ in the linear regime, the shift $\Delta{y}$/$\delta{y}$ must in fact be an exact number of bins, and must be as high as roughly 10. But $\Delta{y}$ depends on $r$, and we want to be able to run multiple simulations with variable $r$, and in non-linear simulations $r$ will be constantly modified (see section~\ref{sec-dsa-kino-nonlinear}) so that we will never know its final value beforehand. To solve this problem we proceed as follows. We use the same momentum resolution $\delta{y}$ to run all shocks, fixed so that $\delta{y} < \Delta{y}(r_{min}) / J$ where $J$ is a chosen integer $\gg1$ and $r_{\rm min}$ is the lowest allowed compression ratio (note that in the case of modified shocks the relevant ratio for decompression is the total one, which will always be greater than the ratio imposed initially, so that $r_{\rm min}$ is well defined). At the end of each shock we measure the actual value $r$ of the compression ratio, re-bin the numerical spectrum $f$ with new resolution $\delta{y'}=\Delta{y}(r)/J$, shift this under-sampled $f$ by $\Delta{y}(r)$ that is by exactly $J$ bins, and then re-bin $f$ back to the nominal resolution $\delta{y}$.

Regarding escapes due to losses we simply give to the code a cut-off momentum $p_{\rm cut}$ above which the CR spectrum $f$ is set to zero.

\subsection{Parallelization}
\label{sec-multi-para}

Even using AMR running multiple realistic shocks simulations can easily be very demanding in computing time, limiting drastically the possibility to do parameter studies and thus fully explore the mDSA mechanism. We note here that even if the code is 1D in space the inclusion of the full spectrum of particles in each cell makes it actually 2D. Thus we now pay more attention to the momentum dimension.

\subsubsection{Principle}
\label{sec-multi-para-principle}

Confronted by this problem, \cite{Jones2005a} use an interesting "coarse-grained finite momentum volumes" technique to lower the constraints imposed by the $p$ dimension. The basic idea under this approach (first introduced by \cite{Jun1999a} and \cite{Jones1999a}) is simply to lower the numerical resolution $\delta y$ in momentum, but prescribing a power-law spectrum shape to each part of the discretized spectrum in order to keep reasonable accuracy. The numerical spectrum is then no longer a piece-wise constant function but a piece-wise linear function. This technique allows reasonably good estimates of the modified shock evolution with unusually low momentum resolutions (which can be as low as 2-3 bins per decade). However we believe that the adiabatic decompression between multiple shocks wouldn't be handled properly by such low resolutions, as it is typically of only 1/5 of decade and has to be well sampled to get precise results in the linear regime (see section~\ref{sec-multi-msa-num}).

In this paper we adopt an alternative way to lower the $p-$dimension numerical cost, without any compromise regarding momentum resolution, which consists simply of fully exploiting the power of modern super-computers by parallelizing the code. We consider here the paradigm, implemented with MPI, which consists of splitting the grid over many processors, each processor running the same code but on its own data -- which still involves some communications between the processors, most notably to define their boundary conditions.

\subsubsection{Implementation}
\label{sec-multi-para-implem}

We have parallelized our code in the momentum space as this is straightforward to implement (as nothing happens in momentum space but a global advection to higher energies) and perfect load-balances can always be obtained (provided we slightly adjust the resolution $\delta{y}$ so that the number of momentum bins is an exact multiple of the number of processors). Parallelizing in space would be more difficult because of space diffusion, and far less efficient because with AMR realistic problems are always ill-balanced, no matter how we split the grids. However parallelization in momentum suffers from two limitations. First not \emph{all} the code is parallelized but only the parts dealing with CR and the maximal efficiency of the parallelization of a code is always limited by its sequential portions -- but realistic problems are CR-dominated so that it is easy to reach very high ratios of the parallelized over un-parallelized fractions of the code. Secondly, this ratio determines only the \emph{maximum} acceleration achievable through parallelization: in practice the effective efficiency of parallelization is limited by the extra cost induced by inter-processor communications, which eventually leads to a saturation of the efficiency -- but we managed to achieve good scalings as shown below.

\subsection{Test 3: doing multiple shocks}
\label{sec-multi-test}

Here we present the evolution of test~2 when multiple shocks are run. To the best of our knowledge, these are the first direct simulations of time-dependent linear and non-linear multiple DSA.

\subsubsection{Test Design}
\label{sec-multi-test-design}

We start from the design of section~\ref{sec-scales-test-design}, with the same hydrodynamical initial conditions for each shock, but an evolving CR population, as explained in section~\ref{sec-multi-msa-num}. We recall that particles are injected at $\log(p_{\rm inj})=-1$. We want here to study acceleration up to $\log(p)=1$. When multiple shocks are run with CR now present everywhere upstream the code faces harder numerical precision issues at high energies. To fix that first we set a bigger maximum momentum $\log(p_{\rm max})=1.5$ and we set bigger sub-grids ($\Lambda=15$). We use the same space resolution as before: $\lambda=0.3$ at $\log(p_{\rm min})=-1.5$ (that is $\lambda=0.1$ at $\log(p_{\rm inj}^0)=-1$). We use a better momentum resolution: $\delta_y=0.036$ corresponding to exactly 64 bins per decade (which, over 3 decades in $p$, allows perfect load balancing when running the code in parallel on clusters of processors that are powers of 2 up to 64). This nominal resolution is adjusted at the end of each shock to be exactly 10 bins per decompression shift (as explained in section \ref{sec-multi-msa-num}). We consider here that $p_{\rm cut} > p_{\rm max}$ so that CR don't escape between two shocks (section \ref{sec-multi-msa-num}).

\subsubsection{Parallelization efficiency}
\label{sec-multi-test-bench}

\begin{figure}
\includegraphics*{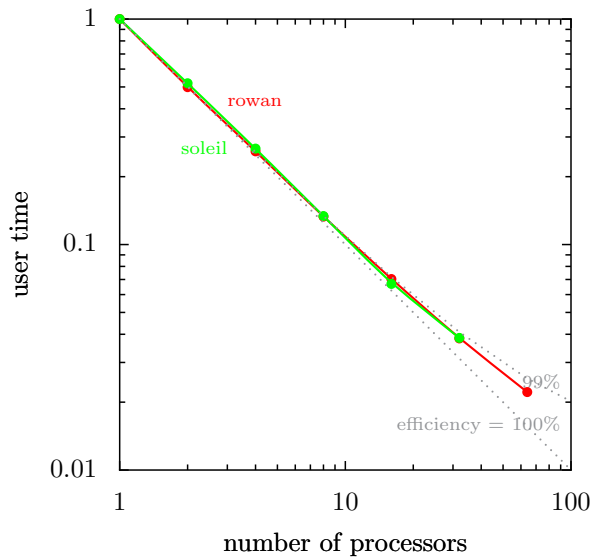}
\caption{User computing time as a function of the number of processors for simulations of section~\ref{sec-multi-test-design}. Measures have been made on two machines, the \emph{soleil} super-computer of the French \emph{Calmip} collaboration (a cluster of 120 Itanium II processors with shared memory) and the \emph{rowan} super-computer of the Irish \emph{Cosmogrid} (a cluster of 256 Xeon processors with Gigabit Ethernet). Tests have been made up to the maximum number of processors available on this middle-class machines (32 on \emph{soleil} and 64 on \emph{rowan}). The computing times have been normalized to emphasize the parallelization scaling with the number of processors.
% (a single Xeon processor is found to be roughly 25\% faster than an Itanium II one). 
The three dotted lines show the theoretical scaling for a perfect parallelization (that is with no induced over-cost) of respectively 99\% and 100\% of the code.}
\label{fig-test3-bench-para}
\end{figure}

Figure~\ref{fig-test3-bench-para} shows the gain brought by parallelization. We obtain good scalings up to a few tens of processors. Thus parallelization allows us to study the acceleration by multiple shocks within the time previously required to study acceleration by a single shock.

In this simulations the DSA operator represents a bit more than 90\% of the computations time, and as the hydrodynamic operator also advects CR the fraction of the code actually benefiting from parallelization is more than 99.5\%.
On slightly less CR-dominated simulations we have observed that the scaling is better on the shared memory machine than on the distributed memory machine, as our code is then bound by communications.
%The scaling is slightly less good on the distributed memory machine than on the shared memory machine, which shows that our code is bound by communications.
%(note that \emph{rowan} has been used here with its standard Gigabit Ethernet network, the experimental Infiniband network gives better scaling).
The slightly less good scaling observed with 64 processors is no surprise given that in that case each processor deals with only 3 momenta cells, which makes a high surface/volume (that is communications/computations) ratio. We note here that a good point of parallelization in momentum is that it's all the more usefull since one wants to investigate high energies. Indeed the higher $p_{\rm max}$, the bigger the momentum grid, and the more processors one can use with a same given efficiency.

\begin{figure}
\includegraphics*{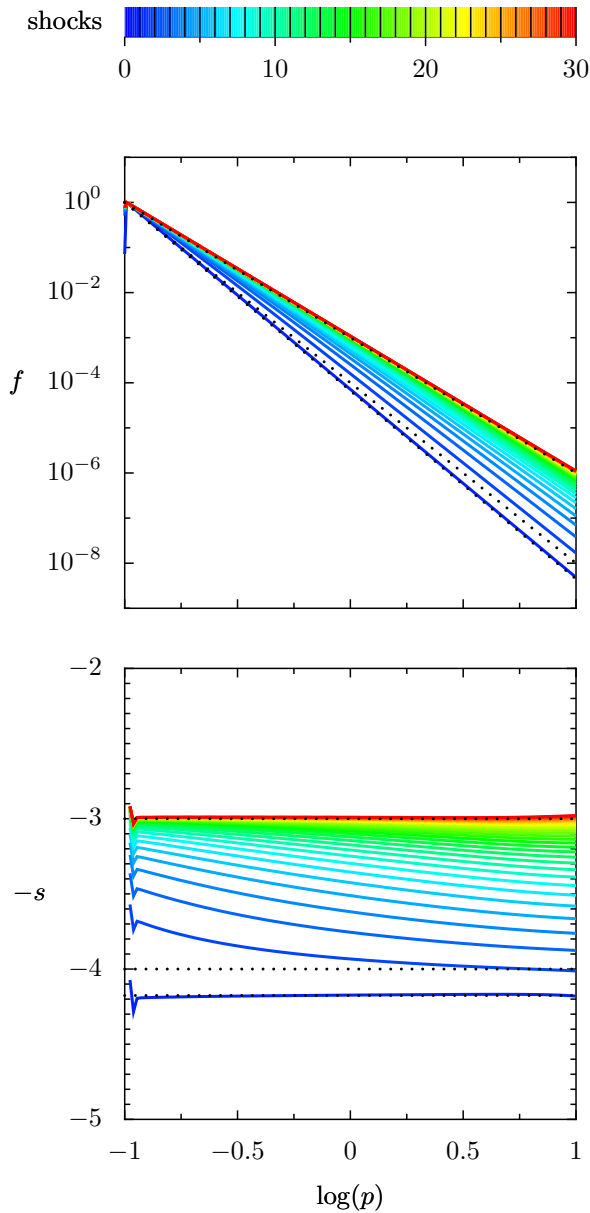}
\caption{Time evolution of the final downstream CR spectra for a sequence of successive linear shocks (see sections~\ref{sec-multi-test-design} and~\ref{sec-multi-test-results-ln} for details). Each coloured line shows the CR distribution just at the end of a shock. The spectra $f(p)$ are shown on top (all normalized so that $f(p_{\rm inj})=1$), where we have added (dotted) the three power-laws of slope $s=3$ and $s=4$ and $s_1=4.18$ the theoretical linear slope for the compression ratio $r=3.55$ of the shocks. These three remarkable slope values are also marked (dotted) on the bottom plot, which shows the spectra logarithmic slopes $-s = \partial\log(f) / \partial\log(p)$. The evolution of the slope from $s_1$ to $3$ with the number of shocks is shown on figure~\ref{fig-test3-multi-t} for three different momenta.}
\label{fig-test3-multi-p}
\end{figure}

\begin{figure}
\includegraphics*{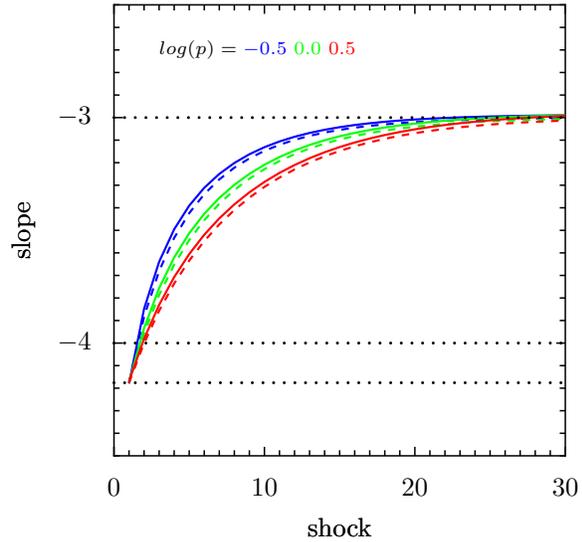}
\caption{Evolution of the final CR spectrum slope at three different momenta ($\log(p)=-0.5, 0.0, +0.5$) for a sequence of successive linear shocks (this figure consists in three vertical cuts in the bottom plot of figure~\ref{fig-test3-multi-p}). The theoretical results (from equation~(\ref{eq-fN})) are added dashed. The three remarkable slopes $s=3$, $s=4$ and $s_1$ are marked dotted.}
\label{fig-test3-multi-t}
\end{figure}

\subsubsection{Linear results}
\label{sec-multi-test-results-ln}

Figures~\ref{fig-test3-multi-p} and~\ref{fig-test3-multi-t} show the evolution of test~2 when 30 such shocks are successively launched with CR back-reaction turned off. In this linear case the CR pressure grows forever, so that we end each shock at the time $t_{\rm end}=300$ corresponding to $\log(p_{\rm max})=1.5$. Figure~\ref{fig-test3-multi-p} shows the evolution of the CR spectra just downstream of the last shock. The spectra are all normalized at the injection momentum to emphasize the slope evolution (the actual normalization rises by a factor of roughly two). We clearly see the convergence of the spectrum from an initial power-law of slope $s=3r/(r-1)$ (the well-known linear solution for a single shock) to a final power-law of slope $s=3$ (the well-known limit in the case of multiple shocks). The slopes plot (bottom) shows that in between the spectrum is never a simple power-law, as the asymptotic convergence to $s=3$ is all the more slow since the momentum is high. The way the spectrum hardens at different momenta is shown on figure~\ref{fig-test3-multi-t}, where we plot the slope as a function of the number $N$ of shocks. Dashed is added the theoretical slope computed (with centered finite differences) from the analytical expression of the spectrum produced after $N$ shocks, which reads (eg \cite{Melrose1993a})
\begin{equation}
\label{eq-fN}
f_N(p) \propto \sum_{i=1}^{N} \frac{s_1^i}{(i-1)!} \left(R^i\frac{p}{p_{\rm inj}}\right)^{-s_1} \left(\ln\left(R^i\frac{p}{p_{\rm inj}}\right)\right)^{i-1}\end{equation}
where $R=r^{1/3}$ is the decompression factor and $s_1=3r/(r-1)$ is the single shock slope. The code reproduces the expected behavior within roughly $1\%$ for the three momenta (close to the maximum momentum the code is less precise). This validation of our code in the linear regime gives us confidence to explore the unknown non-linear regime.

\subsubsection{Non-linear results}
\label{sec-multi-test-results-nl}

\begin{figure}
\includegraphics*{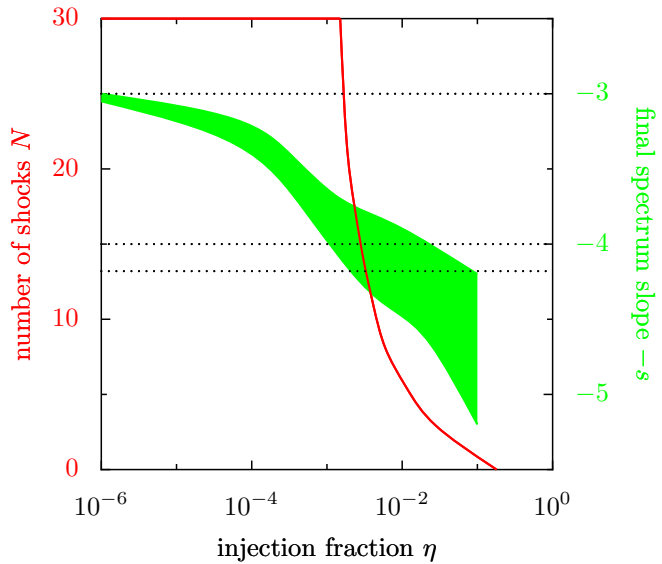}
\caption{Evolution of the number of shocks $N$ (up to 30) that reach a quasi-steady state before complete smoothing and of the range of final CR spectra slopes $-s$ as a function of the injection fraction $\eta$ in the non-linear case (see section~\ref{sec-multi-test-results-nl} for details).}
\label{fig-test3-multi-nl}
\end{figure}

On figure~\ref{fig-test3-multi-nl} we now allow the CR to back-react on the shock. Each shock is run until the downstream CR pressure has converged, before doing decompression and launching the next one. We recall that one of the consequences of CR back-reaction is that $p_{\rm inj}$ varies in time (as the downstream state changes). However we still use here a fixed injection fraction $\eta$, but we run different simulations with different values of $\eta$ as this parameter is quite poorly constrained. To get an overall picture we use a broad range of values of $\eta$, around the value $\eta_0=0.0225$ used since test~1 to match FG parameters. For $\eta \geq 10^{-1}$ (a value which seems physically unreasonably high) the very first shock gets fully smoothed by CR back-reaction before $P_{\rm cr}$ has converged (following FG we consider a shock to be smoothed -- and thus stop injection of fresh CR\footnote{Note that stopping injection doesn't mean stopping acceleration. FG have shown that it is possible to have a CR-dominated "self-sustaining" shock without injection of fresh particles. They had also no advection of particles, whereas with multiple shocks we have to consider the effect of an upstream population too. We postpone the detailed study of this aspect to a future work.} -- as soon as the Mach number of the sub-shock drops below $M_{\rm S,cut}=1.3$). As $\eta$ is lowered to around $10^{-2}$ the first shock can run but the cumulative effect of shocks is such that one shock eventually gets fully smoothed. As $\eta$ is lowered to around $10^{-3}$ the number of shocks~$N$ before smoothing occurs raises exponentially: below roughly $\eta_{\rm c} = 1.5\times10^{-3}$ it seems that virtually any number of shocks could run (although all this shocks are still modified ones). We have limited here the maximum number of shocks to 30, as this seems reasonable for both numerical reasons (given the evolution of the red curve it would take extremely long times to fully explore the very low injection fractions), and physical reasons (considering a few tens of successive strong shocks makes sense in environments such as superbubbles). At the last "complete" shock~$N$ we measure the range of spectra slopes~$s$ (between the injection momentum and the momentum of hardest slope, at which its final decay starts). We observe two evolutions as $\eta$ decreases. First the spectra globally harden as $\eta$ is lowered, which is expected as more shocks can run. Note that as we limit ourselves to $N=30$ the slopes below $\eta_{\rm c}$ can't be directly compared with the slopes above $\eta_{\rm c}$: below $\eta_{\rm c}$ the slopes would get closer to the $s=3$ limit if one would allow for a higher number of shocks. Anyway we see that in the non-linear regime the building of the $s=3$ spectrum within 30 shocks (as on figures~\ref{fig-test3-multi-p} and~\ref{fig-test3-multi-t}) requires an injection fraction lower than $\eta = 10^{-6}$. Second the range of slopes gets constantly narrower, especially below $\eta_{\rm c}$ (that is when when CR no longer limit the number of shocks). Thus this simulations suggest the existence of two regimes of mDSA with respect to the injection ratio $\eta$: there seems to be some critical $\eta_{\rm c}$ (here of roughly $1.5\times10^{-3}$) above which CR dictate the fate of the shocks (producing soft and irregular spectra) and below which CR are almost transparent to the successive shocks (producing harder and more regular spectra). We have observed the same global picture with other simulations (not shown here) involving a constant diffusion coefficient $D$. 

We note that the self-consistent injection fraction proposed by \cite{Blasi2005a} (equation (\ref{eq-eta-blasi})) is here initially $\eta_{\rm B} \approx 10^{-1}$, thus in the regime where CR dominate from the very first shock. This self-consistent $\eta_{\rm B}$ is time-dependent and is lowered as the shock gets modified, but we have observed that the first shock still gets fully smoothed before a quasi-steady state has been reached. Such a very high back-reaction might be surprising for a thermal leakage mechanism. It comes from our particular choice (to match FG parameters) of the ratio of the velocity of injected CR to the downstream sound speed $\xi'=2$, as $\eta_{\rm B}$ has a very strong dependence on this free parameter (recall that $\xi'=1.1\xi$). $\xi'=2$ is a realistic but rather low value, we could suggest as well $\xi'=3$, in which case $\eta_{\rm B}$ is initially $\sim 270$ times lower, that is $\eta_{\rm B} \approx 3.7\times10^{-4}$, that is in the regime where CR are transparent to the successive shocks. Thus this points out that Blasi's model, although providing a self-consistent injection fraction, still requires some initial tuning.

Finally we note that the actual fate of CR-dominated shocks might depend on geometry effects too. For the sake of (both numerical and physical) simplicity we have considered here piston-driven shocks in slab geometry, this work shall now be extended to supernova-like shocks in spherical symmetrical geometry.
We also recall that all this results were obtained with a numerical momentum box limited to $\log(p_{\rm max})=1.5$. Thus their validity depends the assumption that CR accelerated to higher energies: (i) don't have a major impact on the final shock structure (and thus on the spectrum shape at lower energies);
(ii) have enough time to escape from the system between two shocks. Another more fundamental assumption on which we rely (as all other studies of that kind) is that the inertia of the CR is negligible. Although reasonable for sDSA this might be questionable for mDSA because of the cumulative effect of shocks. However we note that the higher the injection fraction, the lower the number of shocks that we actually run. And we have checked that the ratio $\rho_{\rm cr} / \rho$ remains always below 10 \% in all our non-linear simulations.

\section{Conclusion}

We have presented a new code aimed at the simulation of time-dependent non-linear diffusive shock acceleration. It is based on the kinetic approach, coupling the hydrodynamical evolution of the plasma with the diffusive transport of the distribution function of the supra-thermal particles. As such it falls under the legacy of the pioneers (\citealt{Falle1987a}, \citealt{Duffy1992a}) and of the masters (\citealt{Kang1991a}, \citealt{Kang2001a}) of the genre. As the CRASH code it implements an efficient AMR technique to deal with the huge range of space- and time-scales induced by CR diffusion of Bohm-like type. To save even more on computing time we have also parallelized our code in momentum so that we can study acceleration by multiple shocks as fast as acceleration by a single shock. However in many aspects (high-Mach flows, shock tracking, self-consistent injection) our code remains numerically simpler than CRASH -- which can be both a limitation and an advantage. Regarding the physics we note that various mechanisms of importance could be included in the code: self-consistant diffusion coefficient (adding magnetic waves transport), second-order Fermi acceleration (especially between multiple shocks), electrons acceleration (adding radiative losses), CR radiation (hadronic and leptonic)\ldots

We have presented a few tests that show that our code works well, with respect to both the physical accuracy and the numerical efficiency, even in realistic difficult situations. We are now able to investigate in details the various aspects of the DSA mechanism, which 30 years after its early developments still poses some difficulties. In particular we can address the non-linear \emph{multiple} DSA mechanism, which we believe hasn't received so far all the attention it deserves. Our very first results suggest that the injection fraction plays a crucial role. We intend now to study in more details the situation in superbubbles. %(note that we simply have to change the details of the inter-shock physics in order to treat various environments).

\section*{Acknowledgments}

%The authors thank Peter Duffy for helping them with the coupling of hydrodynamic and kinetic aspects and Stephen O'Sullivan for helping them with the coupling of AMR and parallelization.
The authors thank Peter Duffy and Stephen O'Sullivan for helping them with the development of the code presented here.\\
G. F. thanks T. D. and all the staff of the School of Mathematical Science for kindly welcoming him in DCU, Dublin where part of this work has been made.\\
This work has been partly funded by the Egide and Enterprise Ireland \emph{Ulysses} programme, and by the \emph{CosmoGrid} project, funded under the Programme for Research in Third Level Institutions (PRTLI) administered by the Irish Higher Education Authority under the National Development Plan and with partial support from the European Regional Development Fund.\\
The code presented here has been mostly developed on the \emph{soleil} supercomputer of the French \emph{CALMIP} collaboration and on the \emph{rowan} supercomputer of the Irish \emph{Cosmogrid} collaboration.

\bibliographystyle{mn2e}
\bibliography{/Users/gilles/Documents/references}

\label{lastpage}

\end{document}